\documentclass[twocolumn,times,tighten]{aastex63}

\usepackage{graphicx}
\usepackage{amsmath,amstext}
\usepackage{capt-of}
\usepackage{footmisc}
\usepackage{enumitem}
\usepackage{mathrsfs}

\newcommand{\eg}	{e.g.,}%
\newcommand{\ie}	{i.e.,}%
\newcommand{\etal}	{et~al.}%

\newcommand{\HST}	{\emph{HST}}%
\newcommand{\JWST}	{\emph{JWST}}%
\newcommand{\Spitzer}	{\emph{Spitzer}}%

\newcommand{\AV}	{\ensuremath{A_{V}}}%

\newcommand{\betaV}	{\ensuremath{\beta_{V}}}%
\newcommand{\betag}	{\ensuremath{\beta_{g}}}%
\newcommand{\betaVzero}	{\ensuremath{\beta_{V\!,0}}}%
\newcommand{\betalamzero}{\ensuremath{\beta_{\lambda,0}}}%
\newcommand{\SB}	{\code{Starburst99}}%
\newcommand{\BC}	{\code{BC03}}%
\newcommand{\sSB}	{\scode{Starburst99}}%
\newcommand{\sBC}	{\scode{BC03}}%
\newcommand{\SExtractor} {\textsc{SExtractor}}

\newcommand{\nsamp}	{257}

\newcommand{\code}[1]	{\textsf{\small #1}}
\newcommand{\scode}[1]	{\textsf{\scriptsize #1}}

\newlength{\txw}
\newlength{\txh}

\received{January 2, 2019}
\submitjournal{ApJ}
\revised{July 31, 2019}
\accepted{Aug 1, 2019}

\reportnum{astro-ph/1901.00565}

\begin{document}


\title{Analysis of the spatially-resolved $V$$-$3.6\,\micron\ colors and dust
extinction in 257 nearby NGC and IC galaxies}

\author[0000-0001-5120-0158]{Duho Kim}
\author[0000-0003-1268-5230]{Rolf A. Jansen}
\author[0000-0001-8156-6281]{Rogier A. Windhorst}
\author[0000-0003-3329-1337]{Seth H. Cohen}
\author{Tyler Mccabe}

\affiliation{School of Earth \& Space Exploration, Arizona State University,
	Tempe, AZ 85287-1404, USA}


\correspondingauthor{Duho Kim}
\email{Duho.Kim@asu.edu}

\shortauthors{Kim, D., \etal}
\shorttitle{$V$\,$-$\,3.6\,$\mu$\MakeLowercase{m}\ colors within NGC/IC
galaxies}

\begin{abstract} 
We present and analyze spatially-resolved maps for the observed $V$- and
$g$-band to 3.6\,\micron\ flux ratios and the inferred dust extinction values,
\AV, for a sample of 257 nearby NGC and IC galaxies. Flux ratio maps are
constructed using PSF-matched mosaics of SDSS $g$- and $r$-band images and
\Spitzer/IRAC 3.6\,\micron\ mosaics, with all pixels contaminated by foreground
stars or background objects masked out. By applying the \betaV\ method
\citep{T09,T10}, which was recently calibrated as a function of redshift and
morphological type by \citet*{K17}, dust extinction maps were created for each
galaxy. The typical 1-$\sigma$ scatter in \betaV\ around the average, both
within a galaxy and in each morphological type bin, is $\sim$20\%. Combined,
these result in a $\sim$0.4 mag scatter in \AV. \betaV\ becomes insensitive to
small-scale variations in stellar populations once resolution elements subtend
an angle larger than that of a typical giant molecular cloud ($\sim$200\,pc).
We find noticeably redder $V$$-$3.6\,\micron\ colors in the center of
star-forming galaxies and galaxies with a weak AGN. The derived intrinsic
$V$$-$3.6\,\micron\ colors for each Hubble type are generally consistent with
the model predictions of Kim \etal\ (2017). Finally, we discuss the
applicability of the \betaV\ dust-correction method to more distant galaxies,
for which well-matched \HST\ rest-frame visible and \JWST\ rest-frame
$\sim$3.5\micron\ images will become available in the near-future.
\end{abstract}


\section{Introduction}

Cosmic dust between an observer and an astronomical object both absorbs and
scatters incoming light (visible and ultraviolet) from the object, and then
reradiates the absorbed energy in the far-infrared (FIR) \citep[and references
therein]{vandeHulst57, Mathis90}. Our rest-frame ultraviolet (UV)--visible view
of distant galaxies is therefore modulated by dust extinction. The level of
extinction varies across the face of a galaxy and differs from galaxy to
galaxy. As a result, our interpretation of the distribution, evolution, and
properties of stellar populations within a galaxy is significantly influenced
by the intervening dust \citep[\eg][]{Elmegreen80, Kennicutt09}.

Correcting for dust extinction generally requires either costly (or
unattainable) UV to FIR multi-wavelength data in combination with spectral
energy distribution (SED) fitting \citep[and references therein]{Conroy13}, or
a full radiative transfer analysis \citep[and references
therein]{Steinacker13}. In special cases, dust extinction can be corrected in
small areas within a galaxy, provided that suitable tracers exist, such as
hydrogen recombination lines.

\citet[][hereafter T09]{T09} and \citet[][hereafter T10]{T10} developed an
economical method that offered \emph{approximate} dust-extinction correction
called the `\betaV\ method', \betaV\,=\,$f_{V}/f_{L}$, which uses the observed
flux ratios of the visible, $V$-band, and infrared $L$-band, $\sim$3.5\micron,
wavelengths. They applied this method to the surface photometry of the nearby,
late-type spiral galaxy NGC\,959, which unveiled a hidden galactic bar. In
order to place the `\betaV\ method' on a more secure theoretical footing,
\citet*[][hereafter K17]{K17} built a library of spectral energy distributions
(SEDs) by stacking the SEDs of simple stellar populations (SSPs) ---spectral
snapshots of a coeval stellar population--- for large ensembles of stochastic
star-formation histories\footnote{\url{http://lambda.la.asu.edu/betav/}}
(SFHs). The SFHs in K17 are designed to reproduce the mean observed SFH as a
function of morphological class, assuming open-box metallicity evolution.
Furthermore, K17 published both modeler-oriented mass-weighted, and
observer-friendly luminosity-weighted predictions, which were then directly
compared to the luminosity-weighted predictions from integral-field unit (IFU)
observations from \citet[][hereafter GD15]{GD15} to calibrate the empirical
`\betaV\ method'.

After its long-awaited launch, currently scheduled for 2021, the \emph{James
Webb Space Telescope} \citep[\JWST;][]{JWST} will observe distant galaxies at
near-infrared (NIR; 0.6--5\micron) wavelengths with similar spatial resolution
as extant visible wavelength images obtained with the \emph{Hubble Space
Telescope} (\HST). Combining rest-frame \JWST\ $\sim$3.5\micron\ and rest-frame
\HST\ $V$-band observations would extend the `\betaV\ method' from a redshift
of $z$\,$\simeq$\,0 to redshifts $z$\,$\lesssim$\,2. With suitably chosen
filter pairs and the previously computed intrinsic flux ratios from K17, one
could then approximately correct the surface photometry of large samples of
galaxies for dust extinction.

In the present paper, we aim to provide a large database of $g$- and $V$-band
to 3.6\,\micron\ flux ratios for local galaxies and their corresponding
dust-extinction maps. We will analyze the results of the flux-ratios to
evaluate the improvement, fidelity, and robustness offered by the approximate
extinction corrections, as well as the nature of exceptional cases. We can then
highlight coherent stellar structures previously hidden by dust, in addition to
cases where the SED is non-thermal in nature, which can result from (weak)
active galactic nuclei (AGN). We also investigate the dependence on the
physical resolution of a galaxy by comparing subsets of galaxies at different
distances. We then compare our results with those obtained through
multi-wavelength SED-fitting found in the literature.

This paper is organized as follows. In \S2, we explain how the catalog of
sample galaxies was constructed and which data were selected for use. \S3
describes the preprocessing of the data, the derivation of the intrinsic
\betaV-values for each Hubble type, and the creation of the corresponding \AV\
dust map for each galaxy. In \S4.1--2, we illustrate the relationships between
\betaV-values and various galaxy properties. In \S4.3--4, we show the derived
radial \AV-profiles for each Hubble type and the hidden coherent features found
in the \betaV-map. In \S5, we discuss the applicability of the \betaV-method.
We briefly summarize our findings in \S6.

We adopt the Planck\,2015 \citep{Planck15} cosmology ($H_{0}$ = 67.8
km\,sec$^{-1}$\,Mpc$^{-1}$; $\Omega_{m}$ = 0.308; $\Omega_{\Lambda}$ = 0.692),
and we will use AB magnitudes \citep{Oke74,Oke83} throughout.

\section{Sample Selection and Data}

We selected our sample galaxies from the Revised New General and Index
Catalog\footnote{\url{http://www.klima-luft.de/steinicke/ngcic/ngcic.htm}}
(Steinicke 2018; hereafter S18), since it contained data necessary for our
analysis of well-studied nearby galaxies. Of the 13,226 objects in the catalog,
9,995 objects are classified as galaxies. Out of these 9,995 galaxies, 568 had
available \Spitzer\ Enhanced Imaging Product (SEIP) Super Mosaics
FITS\footnote{Flexible Image Transport System \citep{FITS1,FITS2}} images,
which were observed with the InfraRed Array Camera \citep[IRAC;][]{IRAC}
Channel 1 onboard the \Spitzer\ Space Telescope \citep{Spitzer}. These
observations were taken at a wavelength of 3.6\,\micron\ and have a full-width
at half maximum (FWHM) resolution of $\sim$1.6\arcsec. The galaxies in each
mosaic were both well-resolved (size\,$\gtrsim$\,100$\times$ point spread
function; PSF), and roughly centered in the IRAC field-of-view (FOV) (offset
from the IRAC pointing center by $\lesssim$\,1\arcmin). To avoid systematic
uncertainties resulting from very large optical depths through a galactic disk,
the 568 galaxies were reduced to 410 relatively face-on galaxies with
axis-ratio (b/a) larger than 0.5. Column 10 in Table~\ref{tab:sample} gives for
b/a values for individual galaxies. Finally, the 257 galaxies which also had
$g$- and $r$-band mosaics from the Sloan Digital Sky Survey (SDSS) Data Release
(DR) 12 server\footnote{\label{sdss}\url{https://dr12.sdss.org/mosaics}} were
selected for the final sample. The SDSS mosaics have average FWHM values of
$\sim$1.2\arcsec.

\begin{figure}
\center
\includegraphics[width=0.489\txw]{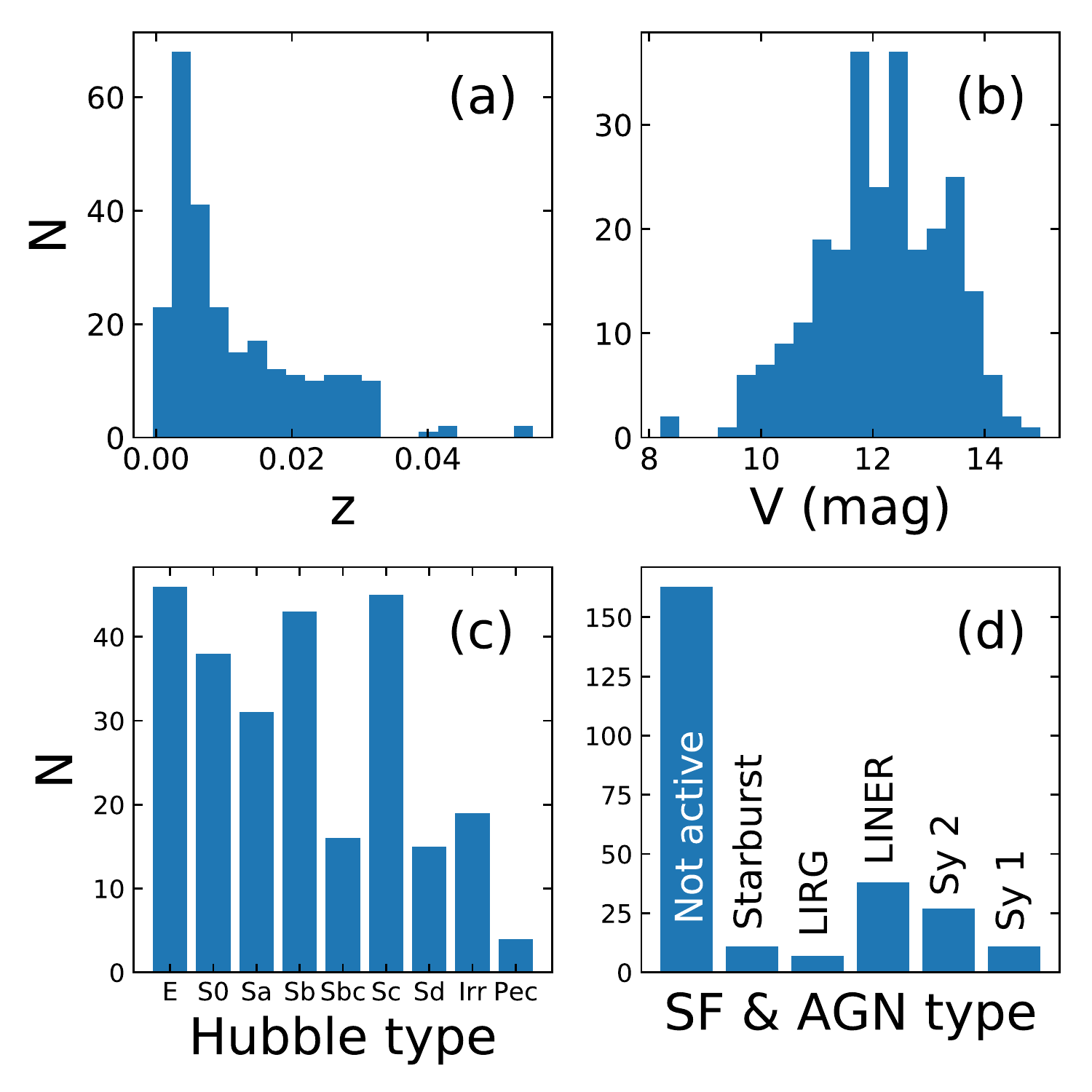}
\caption{\small
Demographics of the selected 257 NGC/IC sample galaxies containing both
\Spitzer/IRAC 3.6\,\micron\ and SDSS $g$ and $r$ mosaics. Distributions detailing
the (a) redshifts, (b) $V$ magnitudes, (c) Hubble type, and (d) SF and AGN
types are shown for the sample galaxies. See \S\ 2.1 and Table~\ref{tab:sample}
for details.\label{fig:demo}}
\end{figure}

Figure~\ref{fig:demo} shows the demographics of the 257 selected galaxies in
our sample. The redshift, $V$-band magnitude, Hubble type, and star formation
and/or nuclear activity type were determined from the NASA/IPAC Extragalactic
Database\footnote{https://ned.ipac.caltech.edu/} (NED), S18, The third
Reference Catalogue of Bright Galaxies \citep[hereafter RC3]{rc3}, and from
NED, respectively.
  %
  %
The Hubble type was determined from the RC3 numeric T type: E\,$=$\
T\,$<$\,$-3$, S0\,$=$\,$-3$\,$\leq$\,T\,$<$\,0, Sa\,$=$\,0\,$\leq$\,T\,$<$\,2,
Sb\,$=$\,2\,$\leq$\,T\,$<$\,4, Sbc\,$=$\,4\,$\leq$\,T\,$<$\,5,
Sc\,$=$\,5\,$\leq$\,T\,$<$\,7, Sd\,$=$\,7\,$\leq$\,T\,$<$\,9,
Irregular\,$=$\,9\,$\leq$\,T\,$<$\,91, and Peculiar\,$=$
\,91\,$\leq$\,T\,$<$\,100. If the `T-type' from RC3 was missing, the value
based on the `Type' from S18 and the `Classification' from NED was assigned.
These are indicated with asterisks in Column 16 of Table~\ref{tab:sample}.
Table~\ref{tab:sample} also lists the celestial (J2000) coordinates, redshifts,
magnitudes ($B$, $V$, and 3.6\,\micron), absolute magnitude (3.6\,\micron),
major-axis size, $r$-band effective radius, Petrosian half-light radius in
$V$\,band, position angle of the semi-major axis, Galactic extinction,
bulge-to-total light ratio, T-type, T-type uncertainty, SF/AGN classification,
and Figure number designation. For the SF and AGN classification, the
`Classification' from NED was used to categorize the galaxies into six types:
`not active', `Starburst', Luminous InfraRed Galaxy `LIRG', Low-Ionization
Nuclear Emission-line Region `LINER', `Seyfert 2', or `Seyfert 1'. If the NED
classification indicates more than one type, we categorize the galaxy with a
preference towards later or more active types. For example, NED classifies
NGC\,0315 as `LINER', `Sy3b', and also `Sy1'. Therefore, we classify NGC\,0315
as `Seyfert 1'. Galaxies without any classification are categorized as `not
active'.

\setlength{\tabcolsep}{2pt}
\begin{deluxetable*}{lccccccccrccccccccccl}
\tabletypesize{\scriptsize}
\tablecolumns{21}
\tablecaption{The list of NGC/IC galaxy sample} 
\tablehead{
\colhead{ID}    & \colhead{R.A.(J2000)}		& \colhead{Decl.(J2000)}		&
\colhead{$z$}		& \colhead{$B$}		& \colhead{$V$}			&
\colhead{3.6\micron} & \colhead{3.6\micron}	& \colhead{$a$}		& \colhead{$b/a$} 	& \colhead{r$_{\mathrm{eff}}$}	& \colhead{$R_{50,V}^{P}$}		& \colhead{PA}		& \colhead{Gal. Ext.}	& \colhead{B/T}		& 
\colhead{T} 		& \colhead{eT}		& \colhead{Type}			& 
\colhead{SF \&}	& \colhead{Classification}	& \colhead{Figure}\\ 
\colhead{}         	& \colhead{(deg)}    & \colhead{(deg)}		&
\colhead{}          & \colhead{(mag)}  	& \colhead{(mag)} 			&
\colhead{(mag)}		& \colhead{(Mag)}	& \colhead{(\arcmin)} 		& \colhead{}			& \colhead{(\arcsec)} & \colhead{(\arcsec)}		& \colhead{(deg)}		& \colhead{(mag)}	& \colhead{}				& 
\colhead{}        	& \colhead{}		& \colhead{}			&	 
\colhead{AGN}        	& \colhead{}			& \colhead{Set 2 \#}\\ 
\colhead{(1)}          & \colhead{(2)}        & \colhead{(3)}       &
\colhead{(4)}          & \colhead{(5)}        & \colhead{(6)}       &
\colhead{(7)}          & \colhead{(8)}        & \colhead{(9)} 		&
\colhead{(10)}          & \colhead{(11)}        & \colhead{(12)} 	&
\colhead{(13)}          & \colhead{(14)}        & \colhead{(15)} 	&
\colhead{(16)}		& \colhead{(17)}	& \colhead{(18)}			&
\colhead{(19)}		& \colhead{(20)}	& \colhead{(21)}
}
\startdata
NGC 0014 &   2.193333 &  15.815556 & 0.002885 & 12.7 & 12.1 & 12.7 & --17.8 & 2.8 & 0.75 &  44.4 & 24.0 & 25.0 & 0.117 & 0.19 & 10.0 & 0.3 & IBm & 0 & (R)IB(s)m pec & 1 \\ 
NGC 0023 &   2.472542 &  25.923778 & 0.015231 & 12.9 & 12.0 & 11.5 & --22.7 & 2.1 & 0.62 &  14.4 &  4.2 & 8.0 & 0.145 & 0.49 & 1.0 & 0.4 & SBa & 2 & SB(s)a;HII;LIRGSbrst & 2 \\ 
NGC 0193 &   9.827458 &   3.331111 & 0.014723 & 13.3 & 12.3 & 11.7 & --22.4 & 1.4 & 0.86 &  19.5 & 13.8 & 55.0 & 0.061 & 0.78 & --2.5 & 0.6 & E/SB0 & 0 & SAB(s)0-: & 3 \\ 
NGC 0266 &  12.449167 &  32.277722 & 0.015547 & 12.5 & 11.6 & 10.3 & --23.9 & 3.0 & 0.97 &  42.3 & 31.2 & 99.0 & 0.217 & 0.28 & 2.0 & 0.7 & SBab & 3 & SB(rs)ab       LINER & 4 \\ 
NGC 0274 &  12.757750 & --7.056944 & 0.005837 & 12.8 & 11.8 & 11.7 & --20.3 & 1.5 & 1.00 & * &  7.8 & 25.0 & 0.211 & 0.45 & --3.0 & 0.4 & E/SB0 & 0 & SAB(r)0- pec & 5 \\ 
NGC 0275 &  12.767500 & --7.066667 & 0.005817 & 13.2 & 12.5 & 12.3 & --19.8 & 1.5 & 0.73 & * & 18.0 & 126.0 & 0.211 & 0.00 & 6.0 & 0.4 & SBcd/P & 0 & SB(rs)cd pec & 6 \\ 
NGC 0309 &  14.177750 & --9.913861 & 0.018886 & 12.5 & 11.9 & 11.8 & --22.9 & 1.94 & 0.69 &  44.4 & 31.2 & 175.0 & 0.156 & 0.06 & 5.0 & 0.3 & SBc & 0 & SAB(r)c        HII & 7 \\ 
NGC 0315 &  14.453667 &  30.352444 & 0.016485 & 12.2 & 11.2 & 10.3 & --24.0 & 2.08 & 0.74 &  36.9 & 23.4 & 43.0 & 0.261 & 0.78 & --4.0 & 0.5 & E2 & 5 & E+:;LINER;Sy3b Sy1 & 8 \\ 
NGC 0337 &  14.958708 & --7.577972 & 0.00549 & 12.1 & 11.6 & 11.0 & --21.0 & 2.9 & 0.62 &  31.5 & 25.8 & 141.0 & 0.316 & 0.31 & 7.0 & 0.3 & SBcd & 0 & SB(s)d         HII & 9 \\ 
NGC 0382 &  16.849458 &  32.403861 & 0.017442 & 14.2 & 13.2 & 12.3 & --22.2 & 0.7 & 1.00 & * &  5.4 & * & 0.173 & 0.61 & --5.0 & 0.5 & E0 & 0 & E:             HII & 10 \\ 
NGC 0383 &  16.854000 &  32.412556 & 0.017005 & 13.2 & 12.2 & 10.8 & --23.6 & 1.6 & 0.87 & * & 15.0 & 30.0 & 0.173 & 0.75 & --3.0 & 0.3 & E-S0 & 5 & SA0-:;BrClG;Sy LERG & 11 \\ 
NGC 0410 &  17.745417 &  33.151889 & 0.017659 & 12.5 & 11.5 & 11.0 & --23.5 & 2.4 & 0.54 &  34.5 & 16.2 & 30.0 & 0.189 & 0.55 & --4.0 & 0.6 & cD & 3 & E+:;LINER      HII & 12 \\ 
NGC 0428 &  18.232125 &   0.981556 & 0.003843 & 11.9 & 11.5 & 12.2 & --19.0 & 4.1 & 0.76 &  49.8 & 34.2 & 120.0 & 0.025 & 0.00 & 9.0 & 0.3 & SBm & 0 & SAB(s)m        HII & 13 \\ 
NGC 0507 &  20.916292 &  33.256056 & 0.016458 & 12.2 & 11.2 & 10.5 & --23.9 & 3.1 & 1.00 &  53.4 & 37.8 & * & 0.165 & 0.39 & --2.0 & 0.3 & E-S0 & 0 & SA(r)0\string^0\string^;BrClG & 14 \\ 
NGC 0514 &  21.016250 &  12.917389 & 0.008246 & 12.2 & 11.7 & 11.7 & --21.1 & 3.5 & 0.80 &  65.7 & 36.0 & 110.0 & 0.073 & 0.03 & 5.0 & 0.3 & SBc & 0 & SAB(rs)c & 15 \\ 
NGC 0596 &  23.217000 & --7.031833 & 0.006258 & 11.8 & 10.9 & 10.6 & --21.7 & 3.2 & 0.66 &  27.3 & 16.8 & 40.0 & 0.107 & 0.62 & --4.0 & 0.5 & E4 & 0 & E+ pec: & 16 \\ 
NGC 0636 &  24.777208 & --7.512611 & 0.006204 & 12.4 & 11.5 & 11.0 & --21.2 & 2.8 & 0.75 &  19.5 & 10.8 & 40.0 & 0.119 & 0.61 & --5.0 & 0.3 & E3 & 0 & E3 & 17 \\ 
NGC 0695 &  27.809333 &  22.582361 & 0.032472 & 13.8 & 12.8 & 11.6 & --24.2 & 0.8 & 0.87 & * &  6.0 & 40.0 & 0.297 & 0.68 & --2.0 & 1.9 & S0/P & 2 & S0? pec;LIRG   HII & 18 \\ 
NGC 0741 &  29.087625 &   5.628944 & 0.018549 & 12.2 & 11.1 & 10.8 & --23.8 & 3.0 & 0.97 &  52.2 & 25.2 & 90.0 & 0.133 & 0.81 & --5.0 & 0.4 & E0 & 0 & E0: & 19 \\ 
NGC 0772 &  29.831583 &  19.007528 & 0.008246 & 11.1 & 10.3 &  9.5 & --23.4 & 7.2 & 0.60 &  77.1 & 43.2 & 130.0 & 0.157 & 0.24 & 3.0 & 0.3 & Sb & 0 & SA(s)b         HII & 20 \\ 
NGC 0777 &  30.062083 &  31.429583 & 0.016728 & 12.5 & 11.5 & 11.0 & --23.4 & 2.5 & 0.80 &  34.5 & 11.4 & 155.0 & 0.145 & 0.66 & --5.0 & 0.3 & E1 & 3 & E1          Sy;LINER & 21 \\ 
NGC 0788 &  30.276875 & --6.815528 & 0.013603 & 13.0 & 12.1 & 11.6 & --22.3 & 1.9 & 0.74 &  17.7 & 12.0 & 111.0 & 0.067 & 0.21 & -.0 & 0.6 & S0-a & 5 & SA(s)0/a:;Sy1  Sy2 & 22 \\ 
IC 0195 &  30.935875 &  14.709278 & 0.012168 & 14.0 & 13.0 & 12.7 & --21.0 & 1.17 & 0.51 & * &  3.6 & 126.0 & 0.137 & 0.72 & --2.0 & 0.8 & S0 & 0 & SAB0\string^0\string^ & 23 \\ 
IC 0208 &  32.115583 &   6.394917 & 0.011755 & 14.2 & 13.4 & 12.9 & --20.7 & 1.8 & 1.00 & * & 19.8 & * & 0.133 & 0.03 & 4.0 & 0.8 & Sbc & 0 & SAbc & 24 \\ 
IC 0214 &  33.523292 &   5.173250 & 0.030224 & 14.7 & 14.2 & 11.3 & --24.4 & 0.8 & 0.75 & * &  3.0 & 162.0 & 0.109 & 0.18 & 7.0* & * & Sd & 2 & I?             LIRG & 25 \\ 
NGC 0985 &  38.657375 & --8.787611 & 0.043143 & 14.0 & 13.4 & 12.8 & --23.6 & 1.0 & 0.90 &  13.2 &  8.4 & 69.0 & 0.093 & 0.73 & 10.0 & 0.8 & Ring/P & 5 & SBbc? p (Ring) Sy1 & 26 \\ 
NGC 0992 &  39.356208 &  21.100833 & 0.013813 & 13.7 & 12.8 & 11.6 & --22.3 & 0.9 & 0.78 & * &  5.4 & 7.0 & 0.409 & 0.45 & 5.0* & * & Sc & 2 & S?             LIRG & 27 \\ 
NGC 1016 &  39.581500 &   2.119250 & 0.022209 & 12.6 & 11.6 & 11.0 & --24.0 & 2.4 & 1.00 &  36.0 & 19.2 & * & 0.053 & 0.69 & --4.5 & 0.5 & E0 & 0 & E & 28 \\ 
NGC 1143 &  43.790458 & --0.177861 & 0.028216 & 14.1 & 13.1 & 12.5 & --23.0 & 0.9 & 0.89 & * &  8.4 & 110.0 & 0.177 & 0.29 & --4.3 & 0.6 & Ring A & 0 & S0 pec (Ring A) & 29 \\ 
NGC 1144 &  43.800833 & --0.183556 & 0.028847 & 13.8 & 13.0 & 12.0 & --23.6 & 1.1 & 0.64 &  16.8 &  9.0 & 130.0 & 0.177 & 0.40 & 3.0* & * & Ring B & 4 & S pec (Ring B) Sy2 & 30 \\ 
NGC 1265 &  49.565250 &  41.857750 & 0.025137 & 14.4 & 13.4 & 11.3 & --24.0 & 1.8 & 0.89 &  41.4 & 12.0 & 125.0 & 0.753 & 0.66 & --4.0 & 0.5 & E5 & 0 & E+             LERG & 31 \\ 
NGC 1275 &  49.950667 &  41.511694 & 0.017559 & 12.6 & 11.9 & 10.8 & --23.7 & 2.2 & 0.77 &  16.8 & 22.2 & 110.0 & 0.697 & 0.29 & 99.0 & * & S0/P & 4 & cD;pec;NLRG;Sy2;LEG & 32 \\ 
NGC 1667 &  72.154750 & --6.319972 & 0.015167 & 12.8 & 12.1 & 11.4 & --22.8 & 1.8 & 0.78 &  18.6 & 12.6 & 20.0 & 0.221 & 0.05 & 5.0 & 0.4 & SBc & 4 & SAB(r)c        Sy2 & 33 \\ 
NGC 1700 &  74.234625 & --4.865778 & 0.012972 & 12.2 & 11.2 & 10.5 & --23.3 & 3.3 & 0.64 &  18.6 &  9.0 & 120.0 & 0.113 & 0.82 & --5.0 & 0.3 & E4 & 0 & E4 & 34 \\ 
NGC 2403 & 114.214167 &  65.602556 & 0.000445 & 8.9 & 8.5 &  8.6 & --17.9 & 21.9 & 0.56 & 143.7 & 126.0 & 127.0 & 0.157 & 0.00 & 6.0 & 0.3 & SBc & 3 & SAB(s)cd;HII   LINER & 35 \\ 
NGC 2415 & 114.236208 &  35.241972 & 0.012622 & 12.8 & 12.4 & 11.6 & --22.2 & 0.9 & 1.00 & * &  7.2 & * & 0.161 & 0.37 & 10.0 & 1.7 & Im & 0 & Im? & 36 \\ 
IC 0486 & 120.087417 &  26.613528 & 0.026875 & 14.6 & 13.7 & 12.6 & --22.8 & 1.11 & 0.66 & * &  6.0 & 139.0 & 0.049 & 0.59 & 1.0 & 0.9 & SBa & 5 & Sy1 & 37 \\ 
NGC 2500 & 120.471708 &  50.737111 & 0.001681 & 12.2 & 11.6 & 11.1 & --18.2 & 2.9 & 0.90 & * & 40.8 & 48.0 & 0.129 & 0.21 & 7.0 & 0.3 & SBcd & 0 & SB(rs)d        HII & 38 \\ 
NGC 2512 & 120.782708 &  23.391833 & 0.015684 & 13.9 & 13.1 & 12.3 & --21.9 & 1.4 & 0.64 &  12.9 & 11.4 & 113.0 & 0.161 & 0.31 & 3.0 & 0.8 & SBb & 1 & SBb            Sbrst & 39 \\ 
IC 2239 & 123.528292 &  23.866361 & 0.020174 & 14.6 & 13.6 & 12.9 & --21.9 & 1.3 & 0.85 & * &  4.8 & 168.0 & 0.185 & 0.46 & 0.0* & * & S0 & 0 & S0? & 40 \\ 
NGC 2552 & 124.835542 &  50.009639 & 0.001748 & 12.6 & 12.1 & 13.1 & --16.4 & 3.5 & 0.66 &  68.7 & 43.2 & 57.0 & 0.165 & 0.00 & 9.0 & 0.3 & SBm & 0 & SA(s)m? & 41 \\ 
\enddata

\tablecomments{Our sample of 257 galaxies, sorted by R.A. Unless stated
otherwise, all table entries are taken or derived from NASA/IPAC Extragalactic
Database (NED)\tablenotemark{a}\\
Table 1 is published in its entirety in the online journal. A portion is shown
here for guidance regarding its form and content. \\
(1)\; Same as S18 \\
(2)\; R.A. in degrees and in epoch J2000.0 \\
(3)\; Decl. in degrees and in epoch J2000.0 \\
(4)\; Redshift from NED\\
(5)\; Total $B$-band magnitude from Steinicke 2018 \\
(6)\; Total $V$-band magnitude from Steinicke 2018 \\
(7)\; Total 3.6\,\micron\ magnitude measured using GALFIT (AB mag; see Appendix C for more information)\\
(8)\; (7) + Distance Modulus from (4): not taking into account K-correction \\
(9)\; Major axis in arcminutes \\
(10)\; Axis-ratio \\
(11)\; Effective radius in $r$-band in arcseconds from RC3, but recomputed by
RAJ from log(Ae) after accounting for the fact that the listed Ae values
were diameters, not radii \\
(12)\; Petrosian Half-light radius measured in $V$-band in arcseconds \\
(13)\; Position angle from S18 from North through East\\
(14)\; Foreground Galactic extinction \\
(15)\; Bulge-to-total light ratio measured in 3.6\,\micron\ using GALFIT \\
(16)\; T-type from RC3 or assigned (with `*') based on (18) and (20) \\
(17)\; T-type error from RC3 \\
(18)\; Morphological type from S18 \\
(19)\; Assigned SF/AGN classification based on (20) [0\,=\,Not active, 1\,=\,Starburst, 2\,=\,LIRG, 3\,=\,LINER, 4\,=\,Seyfert 2, 5\,=\,Seyfert 1] \\
(20)\; Morphological and activity classification from NED\\
(21)\; Figure Set number\\
\tablenotetext{a}{The NASA/IPAC Extragalactic Database (NED) is operated by the Jet Propulsion Laboratory, California Institute of Technology, under contract with the National Aeronautics and Space Administration.}}
\label{tab:sample}
\end{deluxetable*}
\normalsize

For each of the 257 sample galaxies, the IRAC Channel\,1 (3.6\,\micron) SEIP
Cryogenic Release v3.0 Super Mosaics were downloaded from the Spitzer Heritage
Archive\footnote{\url{http://sha.ipac.caltech.edu/applications/Spitzer /SHA/}}
(SHA). The data products called `Mean mosaics' were used for the photometry, and those called the `standard deviation maps' were used for the signal-to-noise ratio (S/N) calculation;
following the method detailed in the SEIP Explanatory Supplement\footnote{\url{https://irsa.ipac.caltech.edu/data/SPITZER/Enhanced/SEIP/docs/seip_explanatory_supplement_v3.pdf}}. For the $V$-band data, mosaic
images of 257 NGC/IC galaxies were downloaded in both the $g$ and $r$ bands
from the SDSS DR12 Science Archive Server\footref{sdss}.
30\arcmin$\times$30\arcmin\ SDSS mosaics with a pixel scale of
0.396\arcsec/pixel were selected, such that they had sufficiently larger FOV
than the SEIP Super Mosaics. All SDSS images covering the area were used, even
if they are not from the \emph{primary}\footnote{
\url{https://www.sdss.org/dr12/help/glossary/\#surveyprimary}} SDSS data set anywhere.

\section{Analysis}

\begin{figure*}
\center
\vspace{-0.1cm}
\parbox[b][0.45\txh][t]{0.03\txw}{\textbf{(a)}}
  \includegraphics[width=0.9\txw]{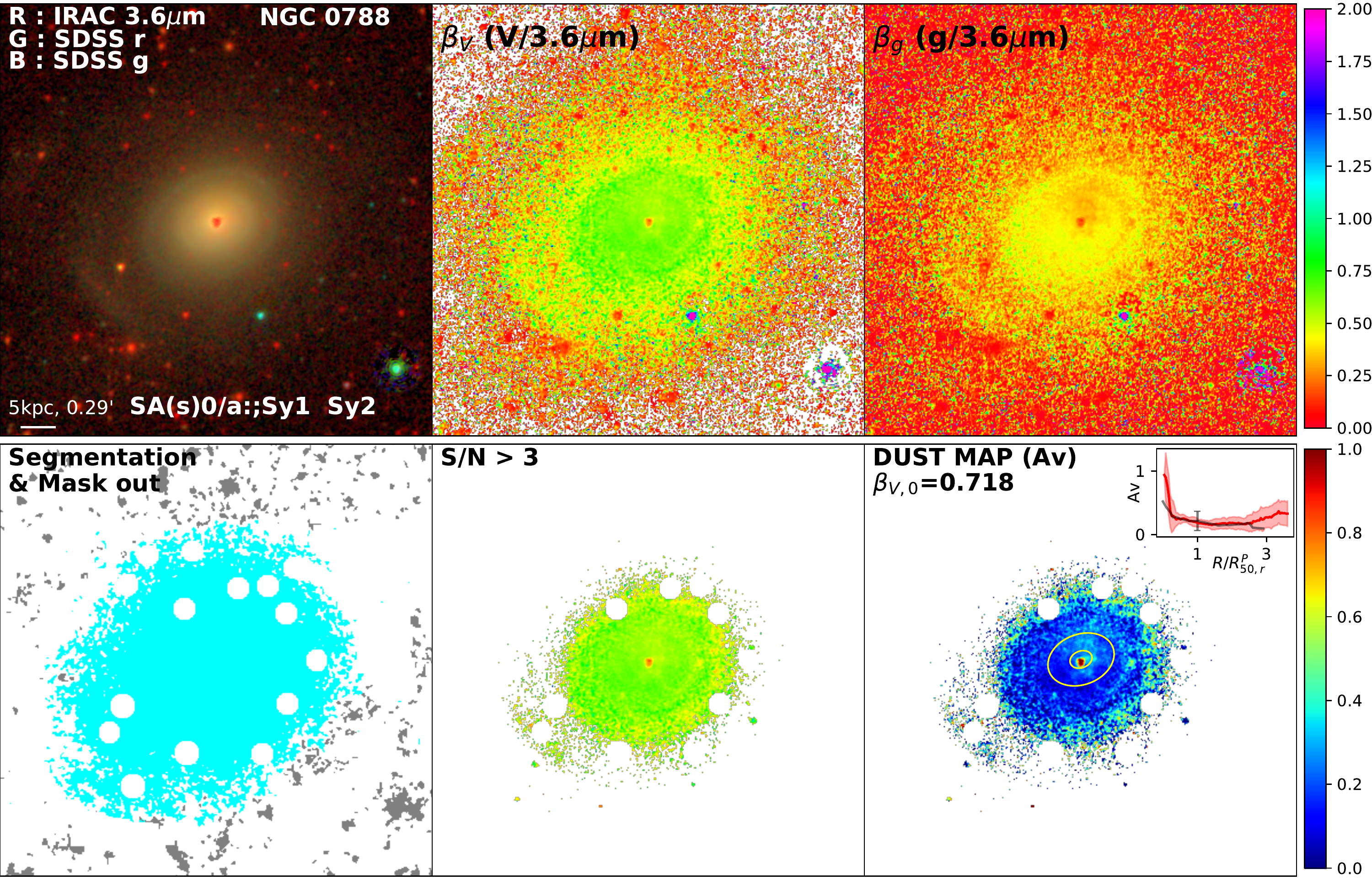} \\
\parbox[b][0.45\txh][t]{0.03\txw}{\textbf{(b)}}
  \includegraphics[width=0.9\txw]{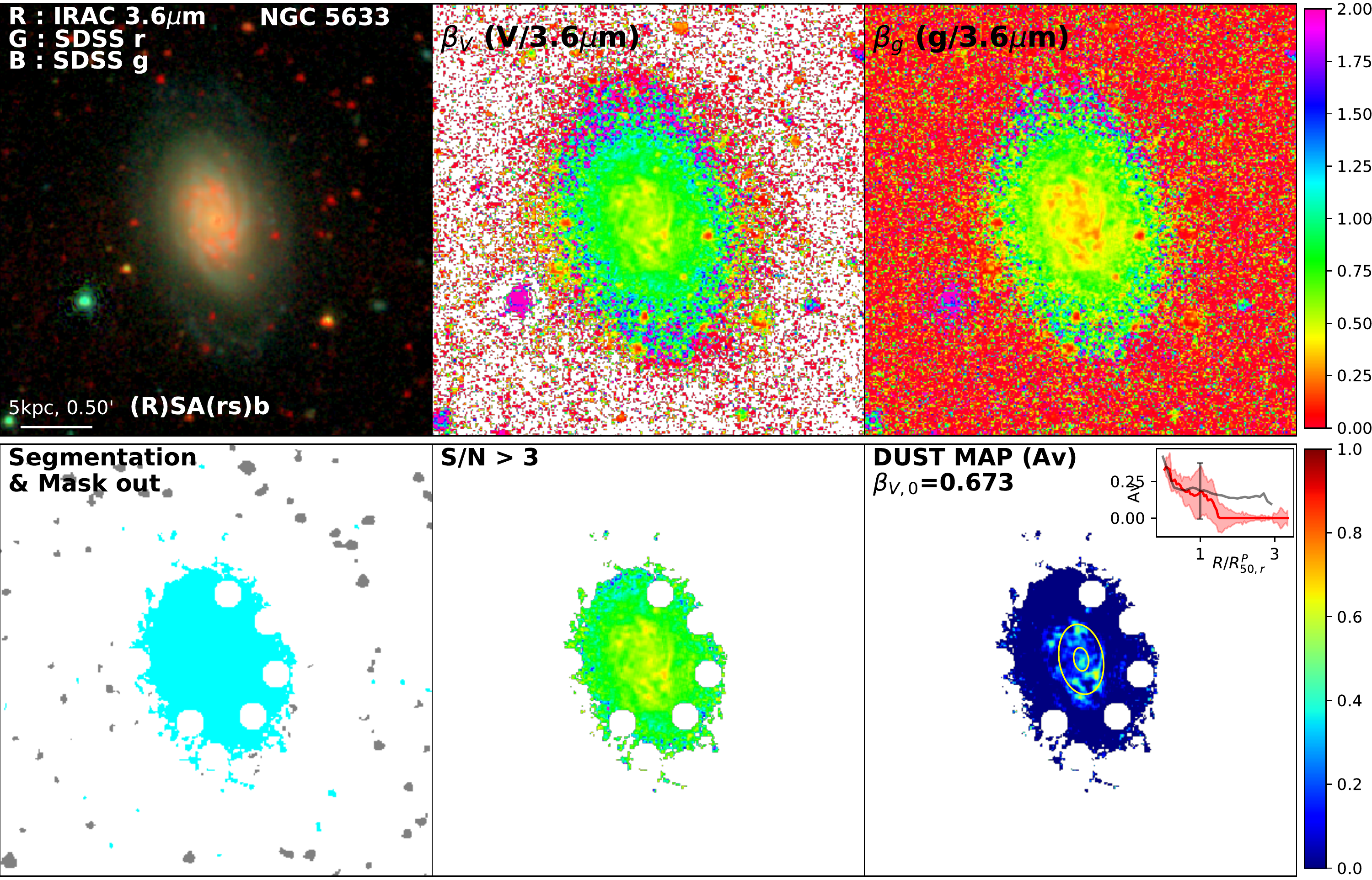}

\caption{\small
\emph{(a)} NGC 0788 and \emph{(b)} NGC 5633 for each: (top left) RGB color
composite from the IRAC 3.6\,\micron, SDSS $r$-, and SDSS $g$-band images, with
the classification from the NASA/IPAC Extragalactic Database (NED) on the
bottom. (top middle) \betaV-image. (top right) \betag-image. (bottom left)
\SExtractor\ segmentation map masking out regions near Galactic stars from the
galaxy of interest (cyan). (bottom middle) \betaV-image of all pixels with a
signal-to-noise ratio greater than three in all mosaics for the region defined
by the bottom-left segmentation map. (bottom right) Dust \AV-map calculated by
the ratio of the \betaV\ to the \betaVzero-values (see Equation~\ref{eq:av}).
The \AV-profile (red) is plotted in the top right corner as a function of
normalized radius with the GD15 \AV-profile for the corresponding Hubble
type (black). The complete figure set (\nsamp\ images) is temporarily available in Google Drive link before the pulication
\footnote{\url{https://drive.google.com/open?id=1gN_fUw_RgEG4MOaGg9extOQ_ZbH2lOPo}}.\label{fig:figset2}}
\end{figure*}

For each galaxy, the PSFs of the SDSS $g$- and $r$-band mosaics were matched to
the Spitzer 3.6\,\micron\ mosaic. The \code{PyRAF}
\code{DAOPHOT}\footnote{\url{http://stsdas.stsci.edu/cgi-bin/gethelp.cgi?daophot.men}} library was used to build model PSFs. Then, the tasks
\code{PSFMATCH}\footnote{\url{http://stsdas.stsci.edu/cgi-bin/gethelp.cgi?psfmatch}} and
\code{WREGISTER}\footnote{\url{http://stsdas.stsci.edu/cgi-bin/gethelp.cgi?wregister}} were used to match the SDSS and \Spitzer\ PSFs, and then register the
image relative to the 3.6\,\micron\ \Spitzer\ mosaic (see Appendix A for details on the PSF matching). Color composite images in the R, G, and B channels are made from
registered IRAC 3.6\,\micron, the SDSS $g$-, and $r$-band mosaics, respectively. These are
available in Figure Set~\ref{fig:figset2}.\footnote{See the electronic version
of this paper for the complete set.}
  %

\smallskip
\figsetgrpstart
\figsetgrpnum{2} 
\figsetgrptitle{
\emph{(top six panels)} NGC 0014 \emph{(bottom six panels)} NGC 0023: (top
left) RGB color composite for IRAC 3.6\micron, SDSS $r$, and SDSS $g$,
respectively, with the classification from the NASA/IPAC Extragalactic Database
(NED) on the bottom. (top middle) \betaV\ image. (top right) \betag\ image.
(bottom left) \SExtractor\ segmentation map masked out the region near the
Galactic stars with cyan color as the galactic region we adopt. (bottom middle)
Same as top middle panel but for the pixels included in the bottom left
segmentation map and having signal-to-noise ratio greater than three in all
mosaics used. (bottom right) Dust, \AV, map calculated by comparing \betaV\
values to the \betaVzero\ values adopted based on the morphology and assumed
metallicity values of Z=0.05, 0.02, and 0.008 for E-S0, Sa-Sbc, and Sc-Sd,
respectively.}

\figsetend

\smallskip

\subsection{\betaV\ map and Segmentation}

The \code{PSFMATCH} task convolves the input image with a convolution kernel to
match a PSF of the input image to the corresponding PSF of the reference image
\citep{Phillips95}:\\
\begin{equation}
 k = \mathscr{F} \bigg\{ \frac{\mathscr{F}(\mathrm{PSF}_{\mathrm{ref}})}{\mathscr{F}(\mathrm{PSF}_{\mathrm{inp}})} \bigg\} ,
\end{equation}
where $k$ is the convolution kernel and $\mathscr{F}$ indicates the
Fourier transform. After the PSFs were matched and the images were registered,
we interpolated the $V$-band flux between the $g$- and $r$-band flux using the
transformation formula from \citet{Jester05} for all stars with
($R$-$I$)\,$<$\,1.15 mag:\\
\begin{equation}
V =g - 0.59 \times (g-r) - 0.01 \ [\mathrm{mag}].
\end{equation}
The rms residual of this transformation equation is 0.01 mag \citep{Jester05}. The \betaV\ FITS
images were produced by dividing the \emph{pseudo} $V$-band co-registered FITS
images by the 3.6\,\micron\ FITS images using the \code{PyRAF} task
\code{IMARITH}\footnote{\url{http://stsdas.stsci.edu/cgi-bin/gethelp.cgi?imarith}}. Examples of the resulting images are shown in the top middle panels of
Figure Set~\ref{fig:figset2}. The \betag\ images are displayed next to the \betaV\ image in each Figure for comparison.

The host galaxy's regions were selected using the segmentation map generated by
\SExtractor\ \citep{SExtractor}, which was obtained by setting
`CHECKIMAGE\_TYPE' as `SEGMENTATION'. The \SExtractor\ parameters
`DETECT\_THRESH', `BACK\_TYPE', `BACK\_SIZE', and `DETECT\_NTHRESH' were
carefully controlled to obtain an optimized segmentation map for each galaxy.
Foreground stars and background galaxies were selected by visual
inspection. Point sources, except the ones in the centers of galaxies, were marked. Extended sources with disparate colors and/or incoherent features were also marked. Their coordinates were recorded manually using the \code{PyRAF}
\code{DAOEDIT}\footnote{\url{http://stsdas.stsci.edu/cgi-bin/gethelp.cgi?daoedit}} task, so that they could be removed from each segmentation map. 
For each object removed from the segmentation map, all pixels within a radius of 10 pixels were set to zero. Finally, each segmentation map was individually analyzed
and then edited, if necessary, using \code{segeditor} \citep{Segeditor}. The
resulting segmentation maps are shown in the lower left panels of Figure
Set~\ref{fig:figset2}. The cyan region is the galaxy of interest and the gray
regions correspond to objects other than the galaxy.

Only the pixels with S/N\,$>$\,3 in all $g$-, $r$-band, and 3.6\,\micron\ FITS
images were used for further analysis. Each 3.6\,\micron\ mosaic comes with an
associated standard deviation map, which we used as a noise map. A similar data
product is not available for the $g$- and $r$-band mosaics, so we measured the
standard deviation of the background noise fluctuations and used it as the
noise value of each mosaic. The final \betaV\ and \AV\ maps used for our
analysis are shown in the lower middle and right panels of Figure
Set~\ref{fig:figset2}.

\subsection{\betaVzero\ derivation}

A \betaV-profile of a galaxy was built by taking the median of the
\betaV-values in elliptical annuli with major axes increasing from 1 pixel to
the maximum visible size of a galaxy (`a' in Table~\ref{tab:sample}). This was
done with a linear step size of one pixel, while the axis-ratio and position
angle (PA) were fixed as `b/a' and `PA' in Table~\ref{tab:sample}. The outer
($r$\,$>$\,20\% of `a') annuli with more than half of pixels masked out were
excluded from the \betaV-profile.

The \betaV-profiles were then converted into \AV-profiles using:
\begin{equation}
    \AV (r) = 2.5 \times \log{(\betaVzero / \betaV (r))} \ ,
\label{eq:av}
\end{equation}
where \betaVzero\ is the global intrinsic \betaV-value, assuming negligible
extinction at 3.6\,\micron\ (see Equation~5 in K17).

The \betaVzero-values for each galaxy or Hubble type were derived by
grid-searching for the \betaVzero-value which had the lowest $\chi^{2}$ between
\AV-profiles from Equation~\ref{eq:av} and the average \AV-profiles for each
Hubble type from GD15 (see Figure~17 in GD15):
\begin{equation}
 \chi^2 = \sum_{r=r0}^{r_n} \frac{ (\AV^{\mathrm{\betaV}}(r) - \AV^{GD15}(r))^2}
	{\sigma^{\betaV}(r)^2}	,
\label{eq:chi}
\end{equation}
where $r$ is the linearly increasing galactic radius normalized by the
Petrosian half-light radius \citep[$R_{50}^{P}$;][]{Blanton01,Yasuda01} (see
Appendix B for more details), \AV$^{\mathrm{\betaV}}(r)$ is the individual or
average \AV-profile from Equation~\ref{eq:av} for a galaxy or Hubble type
respectively, \AV$^{\mathrm{GD15}}(r)$ is the average \AV-profile for each
Hubble type from GD15 for face-on (b/a\,$>$\,0.63) galaxies, and
$\sigma^{\betaV}(r)$ is the scatter in \AV$^{\mathrm{\betaV}}(r)$. We
interpolated over \AV$^{\mathrm{\betaV}}(r)$, so that the number of radius bins
($r_n$\,$=$\,15) was the same as in \AV$^{\mathrm{GD15}}(r)$ within a range of
radius of $r$\,$=$\,0--3\,$\times$\,$R_{50,V}^{P}$. For 69 galaxies, the radial
steps were less than the FWHM of the matched PSF (1.6\arcsec), which indicates
that adjacent data points in the \AV$^{\mathrm{\betaV}}(r)$ profiles are
correlated. However, the \betaVzero-values of interest are derived from the
statistical average of multiple galaxies per Hubble type. The median
\betaV-profiles among $\geq$\,5 galaxies are used in Figures~\ref{fig:prof} and
\ref{fig:prof_bt}. Medians and 1-$\sigma$ ranges in the distribution of the
resulting \betaV-values are shown in Figure~\ref{fig:betav0_dist} and
\ref{fig:betav0}. We refer to \S4.3 for more details. We also verified that
changing the value of $r_n$ from 15 to 30 did not significantly change the
\betaVzero-values.

\subsection{\AV\ map using \betaVzero}

\AV-maps were generated using Equation~\ref{eq:av} as a function of each pixel
instead of as a function of radius (see the lower right panels of the Figure
Set~\ref{fig:figset2}; hereafter `DUST MAP'). 

The \betaVzero-value for each galaxy was selected as the value with the lowest
$\chi^2$ in Equation~\ref{eq:chi}. That \betaVzero-value is printed in the top
left corner of each `DUST MAP' panel. The inset in the top right corner of each
`DUST MAP' panel shows \AV$^{\mathrm{\betaV}}(r)$ (red) and \AV$^{GD15}(r)$
(black) curves as a function of the normalized radius, where the shading
represents the 1-$\sigma$ scatter in the \betaV-profile. The errorbar at a
radius of $R$\,$=$\,$R_{50,V}^{P}$ in the inset represents the uncertainty in
the \AV-value at the half-light radius for the corresponding Hubble type from
GD15.

Yellow ellipses with major axes values of $R_{50}^{P}$ and
3\,$\times$\,$R_{50}^{P}$, corresponding to the tick marks on the \AV-profile
plot, are overplotted on each `DUST MAP' image.

\section{Results}

\subsection{\betaV\ vs. T-type and $z$}

\begin{figure}
\center
\includegraphics[width=0.489\txw]{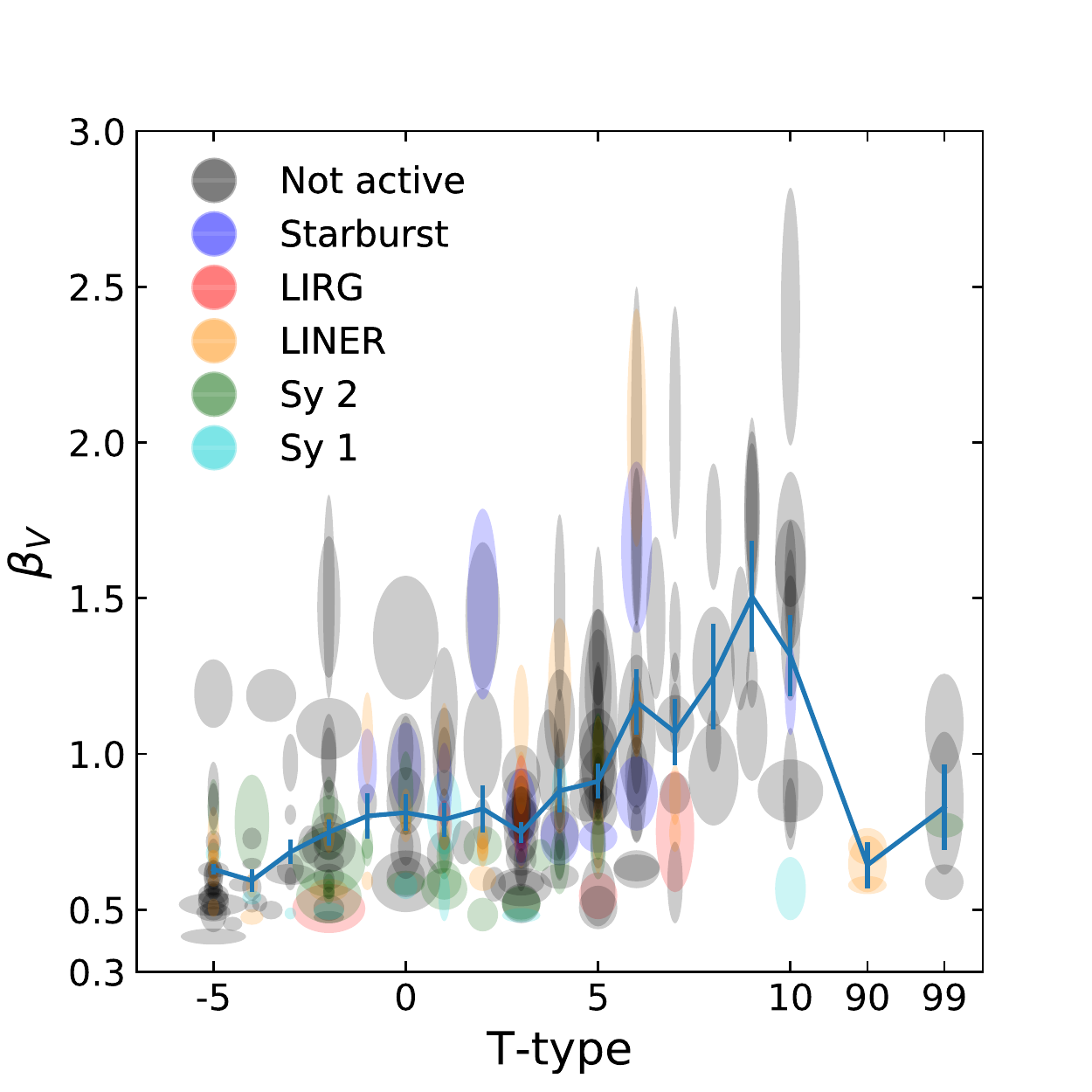} 
\caption{\small
\betaV-values as a function of T-type for our sample galaxies. The colors of
each ellipse represents the SF/AGN type of each galaxy, while the horizontal
and vertical size represents the T-type and the 1-$\sigma$ \betaV\
uncertainties, respectively. The solid line represents the mean and 1-$\sigma$
uncertainty for 1000 randomly generated \betaV-values for
T$-$0.5\,$<$\,T\,$<$\,T$+$0.5. No relationship is seen between the SF/AGN types
and their \betaV-values within each type bin.\label{fig:ttype}}
\end{figure}

The distribution of the \betaV-values for our sample galaxies is shown in
Figure~\ref{fig:ttype} as a function of T-type. The uncertainty in the T-type
and the 1-$\sigma$ error of the \betaV-range are represented by the shape of
the ellipses. Furthermore, SF/AGN types are represented by the color of each
ellipse. For each integer T-type in our sample, a range of T-types was selected
from T$-$0.5\,$<$\,T\,$<$\,T$+$0.5. To investigate the trend in this plot, the
following steps were taken. For each galaxy in every T-type range, a
\betaV-value was sampled from a random distribution of \betaV-values created
using the mean and 1-$\sigma$ error for each galaxy. This was repeated 1,000
times to determine a more statistically significant average \betaV-value for
each T-type. In Figure~\ref{fig:ttype}, this average is shown by the solid blue
line, while the corresponding 1-$\sigma$ error range is illustrated by the
vertical blue lines.
  %
  %

The global \betaV-values (solid blue line) in Figure~\ref{fig:ttype} are
similar to Figure~14 of K17, which corresponds to the range of \betaV-values
derived from the K17 SED models that adopted the observational stellar
population properties from GD15. From both figures (Figure~\ref{fig:ttype} here
and Figure~14 of K17), the average \betaV-values, as well as the scatter of the
values, increases with T-type. The observed 1-$\sigma$ scatter in \betaV\ due
to both the pixel-to-pixel variations within a galaxy, and from galaxy to
galaxy within a Hubble type bin are listed in Table~\ref{tab:scatter}. Within
individual galaxies, the pixel-to-pixel scatter increases toward later Hubble
types, from 0.08 for E to 0.18 for Sb and to 0.33 for Irr\&Pec. From galaxy to
galaxy, the spread within a Hubble type bin also increases from 0.17 for E
through Sbc to 0.45 for Irr\&Pec. On average, the spread in \betaV\ around the
average, both within a galaxy and for each morphological type bin, are
$\sim$20\%, which results in a $\sim$0.4 mag scatter in \AV.
  %
\noindent\begin{table*}
\caption{\small\rule{0pt}{0.4cm}
The observed 1-$\sigma$ scatter in \betaV, \betaVzero, and the calculated
1-$\sigma$ errors in \AV\ for galaxies in each Hubble type
bin.\label{tab:scatter}}
\centering
\setlength{\tabcolsep}{6pt}
\begin{tabular}{ccccccccc}
\hline\\[-16pt]
 & E & S0 & Sa & Sb & Sbc & Sc & Sd & Irr\&Pec\\
 & (46)$^a$ & (38) & (31) & (43) & (16) & (45) & (15) & (23) \\
\hline\\[-14pt]
Within a galaxy$^b$ & 0.08 & 0.15 & 0.20 & 0.18 & 0.24 & 0.30 & 0.31 & 0.33 \\
From galaxy to galaxy$^c$ & 0.15 & 0.19 & 0.16 & 0.19 & 0.17 & 0.27 & 0.32 & 0.45 \\
1-$\sigma$ ranges of \betaVzero$^d$ & 0.15 & 0.25 & 0.38 & 0.38 & 0.35 & 0.63 & 0.45 & \nodata \\
\betaVzero$^e$ & 0.55$^{+0.08}_{-0.07}$ & 0.64$^{+0.13}_{-0.12}$ & 0.72$^{+0.11}_{-0.27}$ &
0.68$^{+0.19}_{-0.19}$ & 0.89$^{+0.15}_{-0.20}$ & 0.96$^{+0.37}_{-0.26}$ &
1.14$^{+0.29}_{-0.16}$ & \nodata \\
Expected $\Delta$\,\AV$^f$ & 0.30 & 0.36 & 0.36 & 0.38 & 0.38 & 0.45 & 0.45 & 0.53 \\
\hline\\[-6pt]
\end{tabular}
\begin{minipage}{\txw}{\small
Notes: \emph{(a)} The number of galaxies in each Hubble type bin. \emph{(b)}
The mean of 1-$\sigma$ scatter in the pixel-to-pixel \betaV-values within each
galaxy (half vertical sizes of the ellipses in Figures~\ref{fig:ttype} and
\ref{fig:betav_z}) in each Hubble type bin. \emph{(c)} The standard deviation
of the galaxy-by-galaxy average \betaV-values (central vertical values of the
ellipses in Figure~\ref{fig:ttype}) in each Hubble type bin. \emph{(d)} The
1-$\sigma$ scatter of the inferred \betaVzero-values for galaxies in each
Hubble type bin. See \S4.3 and Figure~\ref{fig:betav0_dist} for more details.
\emph{(e)} The median and the 1-$\sigma$ ranges of the inferred
\betaVzero-values. \emph{(f)} The calculated 1-$\sigma$ scatters of
\AV-values using the derivatives of Equation~\ref{eq:av}
$\lbrack\frac{dA(x)}{dx}=2.5\times\frac{d\betaV(x)/dx}{ln(10)\betaV(x)}$, where
we adopt (d) for $d\betaV(x)/dx$ and the average \betaV-values for each Hubble
type bin for \betaV(x))$\rbrack$.}
\end{minipage}
\end{table*}
  %
Slowly evolving old and metal-enriched stellar populations in early-type
galaxies tend to be more similar compared to more dynamically evolving young
and low-metallicity galaxies of later types, resulting in a wider \betaV-range
for the latter. However, spiral types with \betaV-values less than 0.5 from
\citet{Brown14} in Figure~14 of K17, are not seen in this sample. This
difference seems to originate from an intrinsic difference in the type of
galaxies from our sample compared to the sample in \citet{Brown14}. Almost all
of \citet{Brown14}'s galaxies have low resolution 5--38\micron\ spectra from
the \Spitzer\ Infrared Spectrograph \citep[IRS;][]{IRS}. This wavelength range
is where the re-radiation from dust clouds dominates. Furthermore, these
differences can also be attributed to the majority of the galaxies in
\citet{Brown14} being of the SF/AGN class, while the majority of our sample
galaxies are listed as `not active'.

\begin{figure}
\center
\includegraphics[width=0.489\txw]{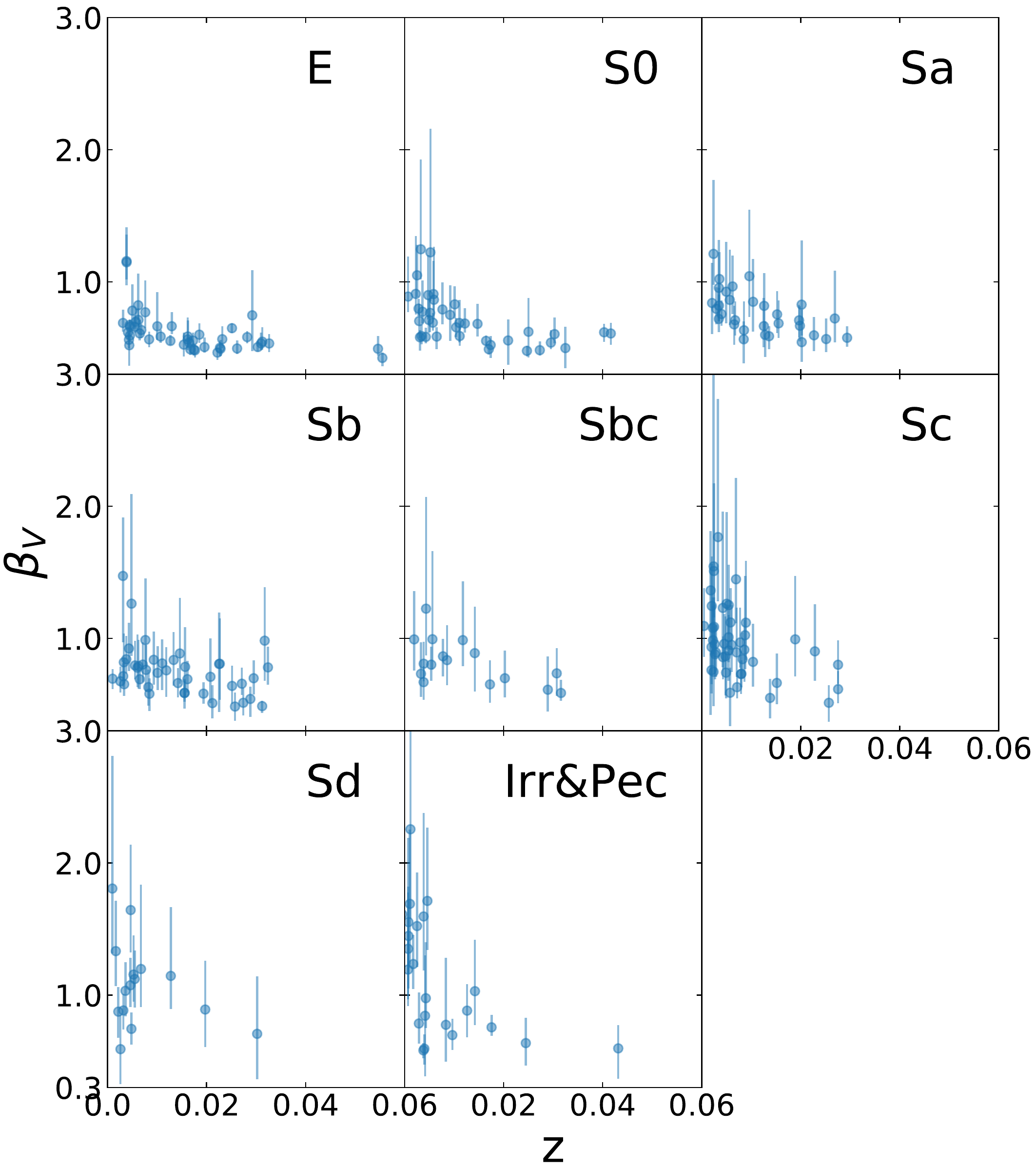}
\caption{\small
The median and 1-$\sigma$ range of \betaV-values as a function of redshift for
each Hubble type in our sample. The scatter in \betaV-values for a single
galaxy and the width of the distribution for a specific Hubble type decreases
as redshift increases, which can be understood as a spatial resolution effect.
The trend seems to level out at $z$\,$\gtrsim$\,0.02, which corresponds to the
spatial resolution achieved by \HST\ and \JWST\ at a redshift of
$z$\,$\simeq$\,0.2.\label{fig:betav_z}}
\end{figure}

The registered image data has a pixel scale of 0.6\arcsec, which is $\sim$38\%
of an FWHM of the \Spitzer/IRAC 3.6\,\micron\ mosaics. This corresponds to a
galaxy surface area of 13$^2$, 250$^2$, 490$^2$, and 718$^2$ pc$^2$ at
$z$\,$=$\,0.001, 0.02, 0.04, and 0.06, respectively. Statistically, as redshift
increases, the sample surface area increases by factors of $\sim$370
($z$\,$=$\,0.02), 1420 ($z$\,$=$\,0.04), and 3050 ($z$\,$=$\,0.06) times
compared to a galaxy at a redshift of $z$\,$=$\,0.001. This reduces the scatter
of any measurements taken by $\sim$19, 38, and 55 times compared to the
measurement taken at a redshift of $z$\,$=$\,0.001 for a completely random
sample. This trend is illustrated in Figure~\ref{fig:betav_z}, where the median
and the scatter in \betaV-values decreases significantly as the redshift
increases. For redshifts less than $z$\,$\lesssim$\,0.02, this trend holds.
This trend can be interpreted as the \betaV-values arising from stellar
populations not being well-mixed (\ie\ being spatially correlated among
adjacent pixels), as well as spatial resolution effects, which cause less
variance in \betaV\ at $z$\,$\gtrsim$\,0.02.
  %

\subsection{Central \betaV\ vs. SF and AGN activity}

\begin{figure}
\center
\includegraphics[width=0.489\txw]{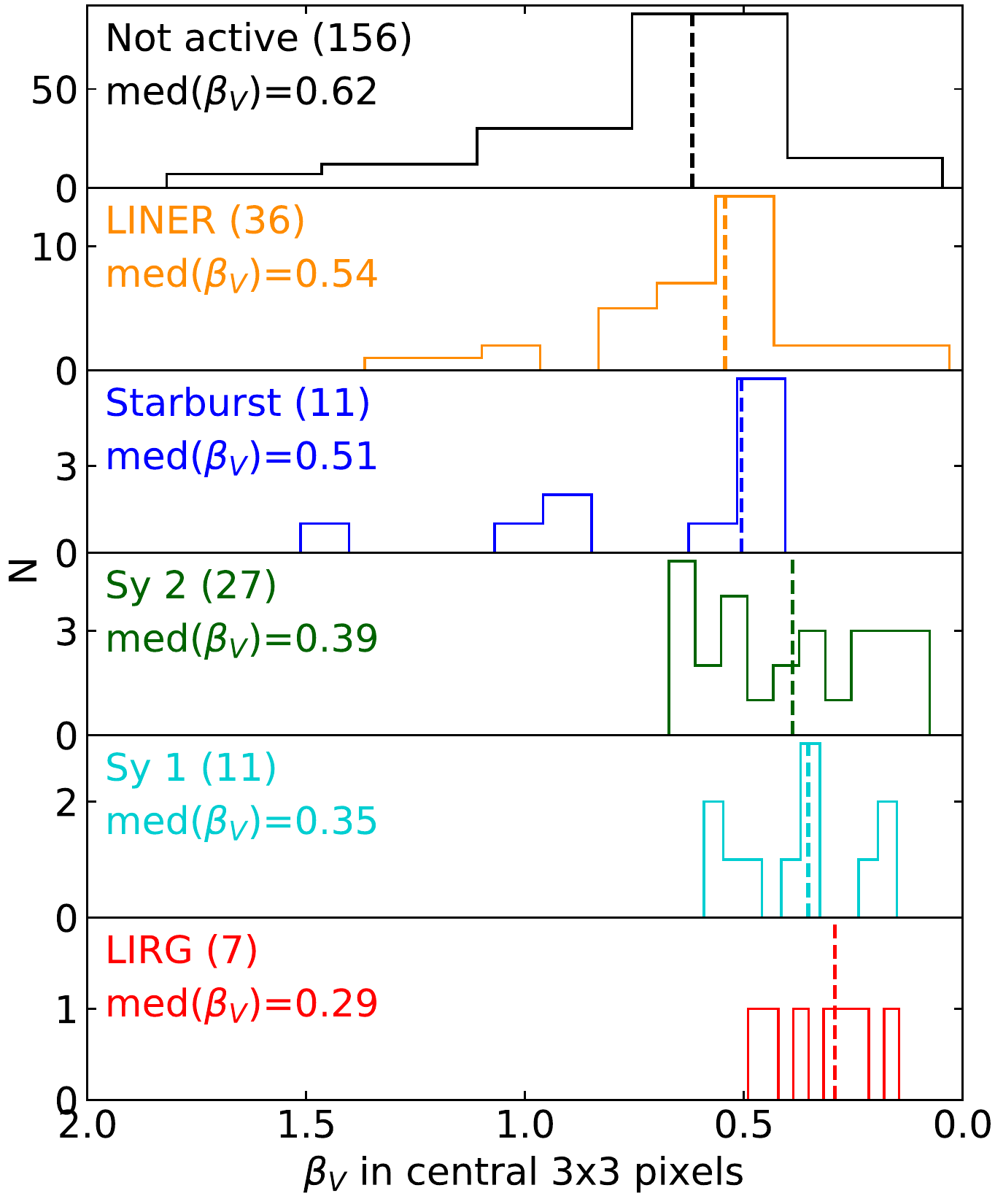}
\caption{\small
Histograms showing the mean \betaV-values for the central 3x3 pixels of each
galaxy. From top to bottom, this is done for 248 NGC/IC galaxies with different
SF/AGN types: Not active, LINER, Starburst, Seyfert type 2, Seyfert type 1, and
LIRG. Compared to average (Not active) galaxies, galaxies with increasing
star-formation and/or nuclear activity have decreasing central \betaV-values,
which implies an increasing amount of dust.\label{fig:agn}}
\end{figure}

In Figure~\ref{fig:ttype}, SF/AGN galaxies are seen to have randomly
distributed global \betaV-values. However, they do have characteristically
lower \betaV-values in their centers. Figure~\ref{fig:agn} shows histograms of
the mean \betaV-values of the central 3$\times$3 pixels for various SF/AGN
galaxies---`Not active', `LINER', `Starburst', `Sy 2', `Sy 1', and
`LIRG'---from top to bottom in order of decreasing median central
\betaV-values. Nine galaxies, whose central 3$\times$3 pixels are masked out,
are excluded.
  %
  %
Central star-forming regions are thought to be enshrouded by molecular clouds,
while accretion disks surrounding supermassive black holes are thought to be
surrounded by a dust torus \citep{Urry95}. Relatively low \betaV-values in the
centers of active galaxies support this idea of dusty environments of active SF
regions and AGN.

\subsection{\betaVzero-values and \AV-profiles}

\begin{figure}
\center
\includegraphics[width=0.489\txw]{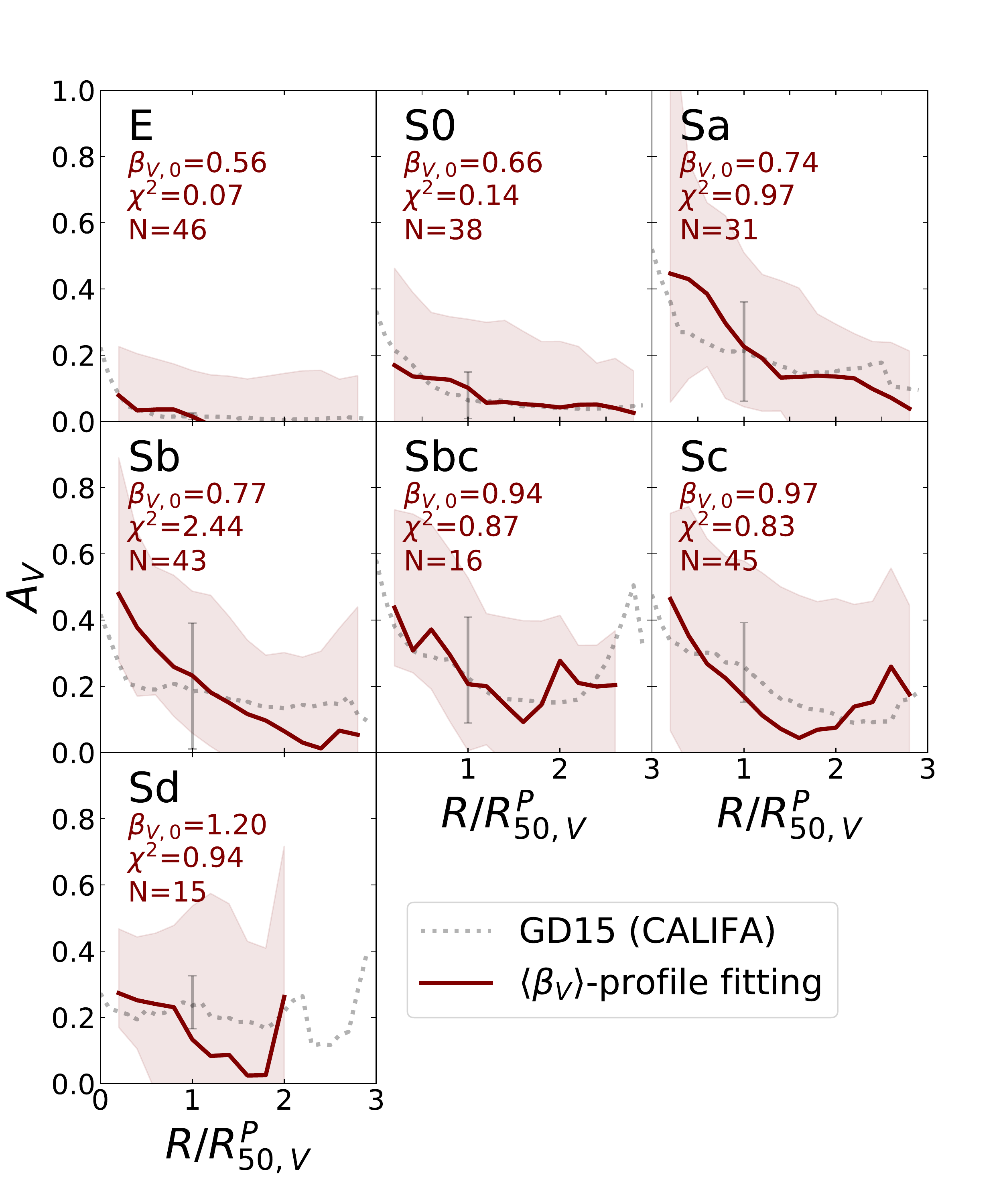}
\caption{\small
\AV-profiles as a function of galactic radius normalized by the Petrosian
half-light radius for different Hubble types. The gray dotted lines are the
averaged \AV-profiles of face-on (b/a\,$>$\,0.63) galaxies as a function of
Hubble type from the CALIFA survey of \citet{GD15}. The error bars at a radius
of $R$\,$=$\,$R_{50}^{P}$ indicate the measurement errors at the half-light
radii for each Hubble type from GD15 (see their Fig.~14), while the solid, dark
brown lines are \AV-profiles generated from Equation~\ref{eq:av} using the
median \betaV-profile for each Hubble type. The \betaVzero-value with the
smallest $\chi^2$ in Equation~\ref{eq:chi} is grid-searched, and the results
are detailed on each panel along with the number of galaxies for each type. The
light purple shaded regions represent the 1-$\sigma$ ranges of the
\betaV-profiles. Sections of the \AV-profiles where the number of available
\betaV-profiles less than five are excluded from the $\chi^2$ fitting. This is
the case for the outskirts of the Sd types. Overall, the profiles agree with
the CALIFA survey results from GD15. Slight deviations are seen in the slopes
of spiral galaxies (Sa, Sb) and in the stochastic fluctuations of galaxies with
small sample sizes (Sbc, Sd).\label{fig:prof}}
\end{figure}

Figure~\ref{fig:prof} shows \AV-profiles for each Hubble type, corresponding to
the least square fitting of the median-combined \betaV-profile to the average
\AV-profile from GD15. The \betaVzero-values and numbers of galaxies for each
Hubble type are also listed. Spiral (Sa, Sb) galaxies have steeper \AV-profiles
than what was calculated in GD15. Sbc and Sd types show larger stochastic
fluctuations, because they have fewer than half of the number of objects
compared to other galaxy type bins. The scatter in the \betaV-profiles results
in larger uncertainties than what was derived from the measurement error at
$R$\,$=$\,$R_{50}^{P}$ from GD15. This is expected, since the \betaV-method
uses only two broadband images and a single color \betaVzero, whereas GD15 used
three-dimensional spectral images and a library of SED models.
  %

\begin{figure}
\center
\includegraphics[width=0.5\txw]{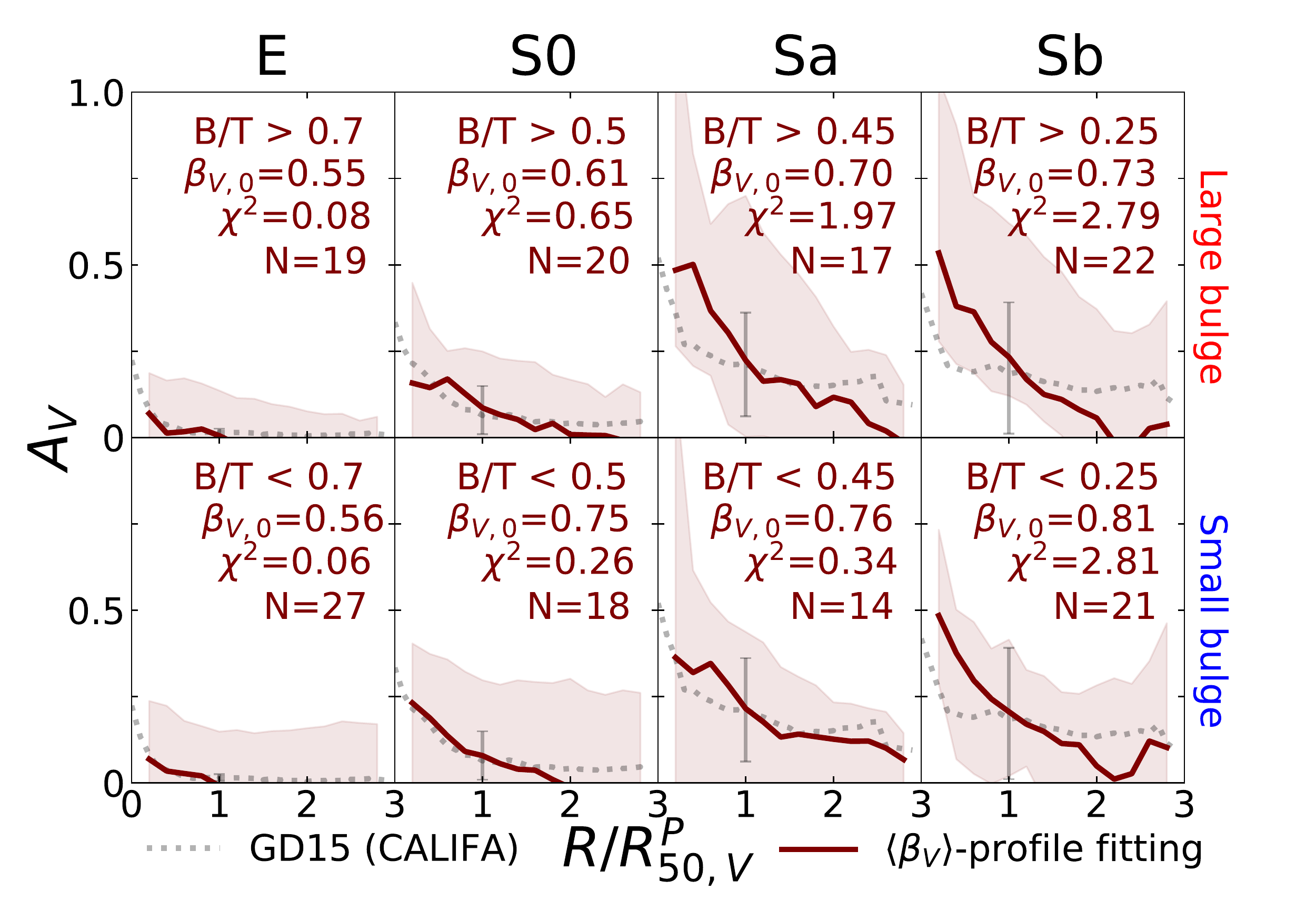}
\caption{\small
Same as Figure~\ref{fig:prof}, but with each Hubble type divided into large and
small bulge subgroups. The subgroups are defined by bulge-to-total light ratios
larger or smaller than 0.7, 0.5, 0.45, and 0.25 for the E, S0, Sa, and Sb
types, respectively. \AV-profiles for Sa type galaxies are in closer agreement
for the small-bulge subgroup, which is a direct result of a more uniform
stellar population within that subgroup.\label{fig:prof_bt}}
\end{figure}

The E--Sb types were subgrouped into `Large bulge' and `Small bulge' categories
by their bulge-to-total light ratios (B/T) derived using \code{GALFIT}
\citep{GALFIT1, GALFIT2} (see Appendix C for details). The B/T threshold values
were arbitrarily chosen, so that there were comparable numbers in each of the
two subgroups. The \AV-profile fitting process was then repeated (see
Figure~\ref{fig:prof_bt}), which significantly reduced the $\chi^2$ for the
small-bulge subgroup of the Sa type. More uniform stellar populations of
galaxies in a subgroup result in more reliable \AV-profiles using the
\betaV-method. This is discussed in \S\ 5.

\begin{figure}
\center
\includegraphics[width=0.5\txw]{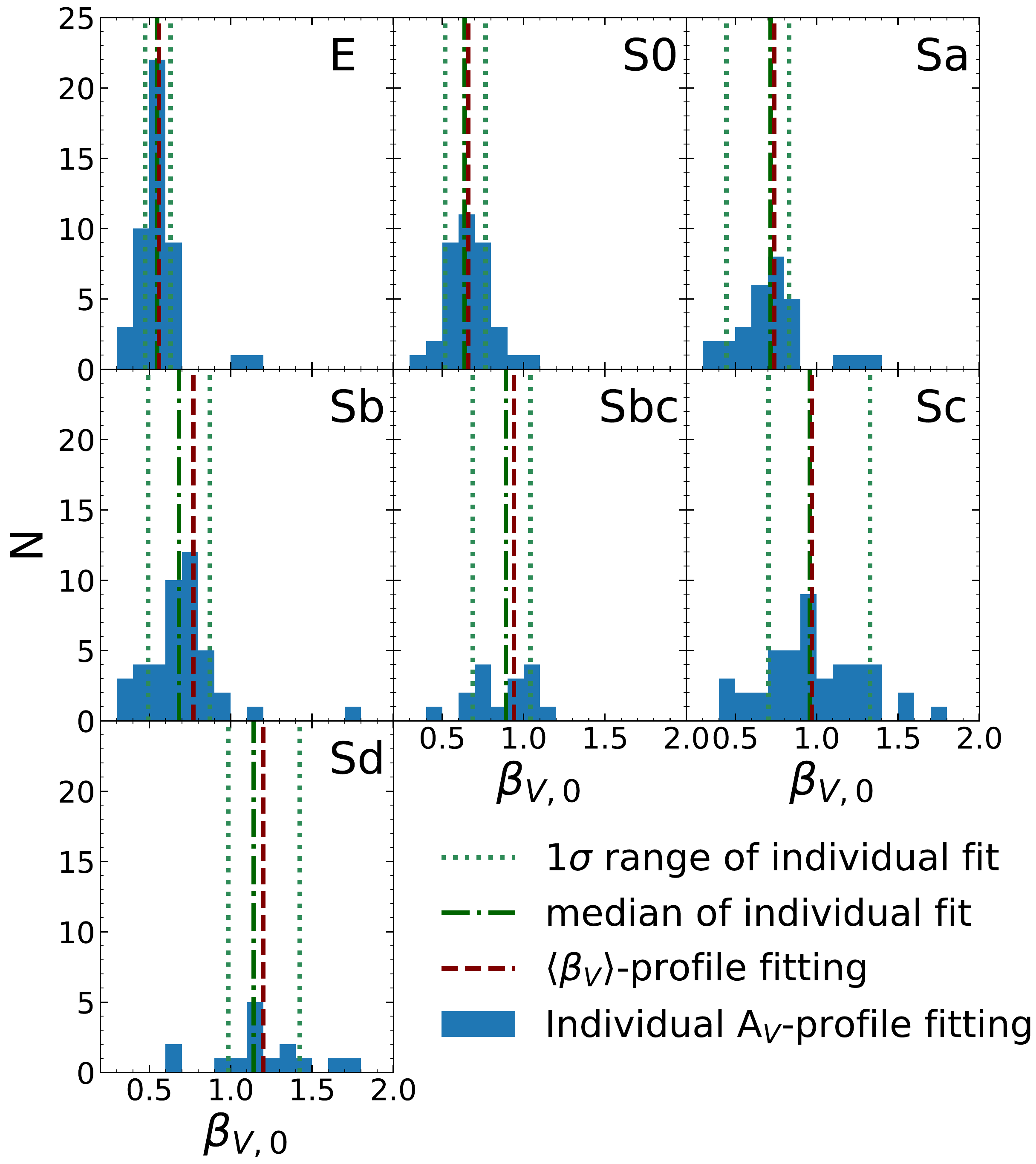}
\caption{\small
Histograms of the \betaVzero-values corresponding to the least square fitting
of the \AV-profiles of each galaxy to the average \AV-profile of each Hubble
type from GD15. The green dot-dashed and dotted lines represent the median and
1-$\sigma$ ranges, respectively. The brown dashed lines show the least square
fitting result of median-combined \betaV-profiles for each Hubble type (see
Figure~\ref{fig:prof}).\label{fig:betav0_dist}}
\end{figure}

  %
Figure~\ref{fig:betav0_dist} shows the histograms of the \betaVzero-values
derived individually for each galaxy. We obtained the \betaVzero-values for
each galaxy by least square fitting its \betaV-profile to the average
\AV-profile of the corresponding Hubble type from GD15. The \betaV-profile is
converted to the \AV-profile with a reference \betaVzero-value (see
Equation~\ref{eq:av}). The median and 1-$\sigma$ ranges are shown as dot-dashed
and dotted green vertical lines, while the \betaVzero-values derived from the
median-combined \betaV-profiles are shown as dark brown dashed lines. The
median values and associated ranges are listed in Table~\ref{tab:scatter}. The
1-$\sigma$ ranges are comparable to the quadratic sum of the pixel-to-pixel and
the galaxy-to-galaxy \betaV-variations in Figure~\ref{fig:ttype}. The median
values derived with these complementary methods agree to within
$\pm$\,1\,$\sigma$.

\subsection{Hidden Coherent Features}

Coherent features buried in a single band or even in composite images (such as
top left panel of Figure Set~\ref{fig:figset2}) stand out in color images, such
as in the \betaV, \betag, and/or \AV-maps. The features found in these color
maps are detailed in Table~\ref{tab:features}, along with notes on their
corresponding galaxies.
  %
\noindent\begin{table*}
\caption{\small\rule{0pt}{0.4cm}
Coherent features revealed in the \betaV, \betag, and/or \AV\ images and lists
of galaxies containing the features\label{tab:features}}
\centering
\setlength{\tabcolsep}{9pt}
\begin{tabular}{ll}
\hline\\[-16pt]
Features (Number of galaxies) & Galaxies (In the order of ID)  \\
\hline\\[-14pt]

Star-forming regions (15) & \textbf{NGC}\,14, 337, 2500, 2552, 2730, 3020, 3381,
3395, 3622, 3794, 3870, 3906, 5668, 5669, 5691 \\[-1pt]
Dust lanes (43) & \textbf{IC}\,676, 1065, 2239, 1551, 2637, \textbf{NGC}\,23,
266, 275, 1667, 2512, 2608, 2712, 2731, 2742, 2750, \\[-3pt]
 & 2776, 2782, 2824, 2893, 2906, 3015, 3049, 3055, 3265, 3349, 3351, 3489,
3655, 3720, 3726, \\[-3pt]
 & 3799, 3801, 3811, 3921, 4194, 5631, 5633, 5992, 7080, 7177, 7479, 7653, 7798
\\[-1pt]
Star-forming regions & \textbf{IC}\,1076, \textbf{NGC}\,275, 309, 337, 2403,
2415, 2731, 2742, 2776, 2906, 3020, 3055, 3184, 3192, \\[-3pt]
and dust lanes (30) & 3212, 3239, 3310 (pec), 3344, 3370, 3381, 3395 (pec), 3486,
3646, 3659, 3686, 3726, 3794, \\[-3pt]
 & 3799, 3811, 3870, 4234, 4420, 4449 (Sbrst), 5660, 5691, 5936, 5992, 7714 (pec),
7741, 7753 \\[-1pt]
LIRG-like thick dusty cloud (18) & \textbf{IC}\,214, 730, 2520 (with blue halo),
\textbf{NGC}\,992, 1144 (pec,Sy2), 3593 (Sy2), 3655, 3683, \\[-3pt]
 & 3822 (Sy2), 3839, 3934, 4207 (elongated), 4414 (LINER), 5541, 5653,
5953, 5990, 6240 \\[-1pt]
Central dust torus (10) & \textbf{IC}\,730, 2551, 2637 (Sy1.5),
\textbf{NGC}\,383 (LERG$^1$), 3015 (NLAGN$^2$), 3349, 3921 (pec), 4014, 4290,
\\[-3pt]
 & 7742 (LINER) \\[-1pt]
Central dust cloud & \textbf{IC}\,486 (Sy1), 676, 691, 3050, 5298 (Sy2),
\textbf{NGC}\,788 (Sy1,2), 985 (Sy1), 1275 (cD;pec;NLRG$^3$; \\[-3pt]
 & Sy2;LEG), 2782 (Sy1, Arp 215), 2824 (Sy?), 2831, 3212, 3561, 3720, 3781,
3839 (with blue disk), \\[-3pt]
 & 3928, 4150, 4162 (AGN), 4194 (pec), 4335, 4369 (intertwined with blue), 4378
(point-like; Sy2), \\[-3pt]
 & 4412 (LINER), 7177 (LINER), 7674 (Sy2) \\[-1pt] 
Star-forming ring & \textbf{NGC}\,3011 \\[-3pt] with outer dust ring (1) & \\[-1pt]
Dust ring (2) & \textbf{NGC}\,2844, 3032 \\[-1pt]
Polar dust ring (2) &
\textbf{NGC}\,2685(Arp\,336)\footnote{\url{https://apod.nasa.gov/apo
d/ap140314.html}}, 3801 \\[-1pt]
Outer blue ring (3) & \textbf{NGC}\,3938, 4162, 7217 \\[-1pt]
Dusty disk in E type (1) & \textbf{NGC}\,3656 (Arp\,155, pec, LINER) \\[-1pt]
Blue disk in E type (2) & \textbf{IC}\,692, \textbf{NGC}\,3011 \\[-1pt]
Central blue region (2) & \textbf{NGC}\,3773 (\ion{H}{2}), 7077 (\ion{H}{2})
\\[-1pt]
M82-like outflow (?) (1) & \textbf{NGC}\,3622 \\[-1pt]
Green center bluer outskirt region & \textbf{IC}\,208, 2239, 3050,
\textbf{NGC}\,274 (pec), 275 (pec), 309, 428, 2043, 2500, 2604, 2730, 2750,
2776, \\[-3pt]
(Negative color gradient) (32) & 2893, 3032, 3049, 3055, 3162, 3319, 3344, 3370,
3489, 3646 (ring), 3659, 3773, 4136, 4234, \\[-3pt]
 & 4412, 4420, 5584, 5585, 7731 \\[-1pt]
Blue center with green outskirt (3) & \textbf{NGC}\,3741 (BCD$^d$), 4068, 6789
\\[-3pt]
(Positive color gradient) & \\

\hline\\[-6pt]
\end{tabular}
\begin{minipage}{\txw}{\small
Notes: (1) Low-excitation radio galaxy; (2) Narrow-line AGN; (3) Narrow
emission-line radio galaxy; (4) Blue compact dwarf; pec: peculiar, Sbrst: Starburst.}
\end{minipage}
\end{table*}

Star-forming regions and dust lanes are the most common features, which have
bluer and redder ($V$-3.6\,\micron) colors, respectively.

LIRG-like galaxies are dominated by optically-thick dust clouds that are likely
to be undergoing cold gas accretion from the cosmic web, harboring an AGN
\citep[see][for \eg\ the case of NGC\,6240]{Muller-Sanchez18,Saito18}, or
experiencing the aftermath of a recent merger \citep[see][for \eg\ the case of
NGC\,4414]{Pingel18}.

AGN appear to have at least local dust clouds such as NGC\,1275
\citep{Tanada18} and NGC\,985 \citep{Appleton93}, except the active, dustless,
bulgeless, intermediate-mass black hole in NGC\,3319 \citep[see][]{Jiang18}.

Inner and outer rings in NGC\,3011 have been found by \citet{GildePaz03}. The
inner ring could have been produced by a starburst-driven shock interacting
with the surrounding medium \citep{Marino13}.

Dust rings in NGC\,2844 and NGC\,3032 are also found and included in the Atlas
of Resonance Rings As Known In S$^4$G \citep[ARRAKIS;][]{Comeron14}. These
rings appear to be formed by a dynamical resonance resulting from
non-axisymmetries in galaxy disks \citep{Comeron14}.

NGC\,2685 (Arp 336, also known as ``The Helix Galaxy'') is a well-known polar
ring galaxy \citep{Sandage61,Eskridge97}, which has an outer ring perpendicular
to the galactic disk. This is seemingly the remains of a merger with a
polar-orbiting smaller galaxy.

NGC\,3801 could be at an earlier stage of an NGC\,2685-like merger
\citep{Hota12}. The merging event $\sim$2--3\,Gyr ago is presumably responsible
for nuclear-ring in NGC\,7742 \citep{Martinsson18}, while the minor merger
$\sim$200\,Myr ago \citep{Knierman12, Knierman13} is responsible for the
central dust in NGC\,2782 and in NGC\,4194 \citep{Konig18}.

\citet{Balcells97} argues that the dusty disk, two tidal tails, and shells in
E-type NGC\,3656 is the outcome of a disk-disk major merger. Conversely, the
blue disk in the E-type, but low-mass, ``blue sequence'' galaxy IC\,692 is from
cold-mode gas accretion \citep{Moffett12}.

Both NGC\,3773 and NGC\,7077 have central blue features that are classified as
\ion{H}{2} regions, which are probably caused by Wolf-Rayet stars
\citep{Miralles-Caballero16}.

The sandglass-shaped feature in NGC\,3622, which resembles the outflow of M82,
has never been reported. SDSS DR12 classifies this galaxy as a starburst
galaxy, which reinforces the idea of an outflow-like feature. Strong H$\alpha$
and \ion{O}{3} emission lines might be the cause of this particular
feature\footnote{\url{http://skyserver.sdss.org/dr12/en/tools/explore/Summary.aspx?id=1237651273508651034}}.

In total, 32 galaxies have noticeable negative color gradients, while 3 have
positive color gradients. The color gradients of NUV(\emph{Galaxy Evolution
Explorer (GALEX)})\,/\,3.6\,\micron\ (S$^4$G) observations for the disks of 1931
nearby galaxies ($z$\,$<$\,0.01) also show negative color gradients \citep[see
Figure~6 in][see also \citealt{deJong96}, \citealt{Taylor05},
\citealt{Kim13}]{Bouquin18}. These color gradients are in agreement with the
global scenario of inside-out formation of disks.

\section{Discussion}

In this study of 257 NGC/IC galaxies at a redshift of $z$\,$\simeq$\,0, we find
that:

\begin{enumerate}[label=\arabic*)]

\item \emph{Early-type galaxies have lower and narrower ranges of \betaV\ and
\betaVzero-values than later type galaxies} (see Figures~\ref{fig:ttype} and
\ref{fig:betav0_dist}). This trend can be explained by an increased amount of
old stellar populations with quenched star-formation in early-type galaxies.

\item \emph{Spiral galaxies have steeper \AV-profiles.} The \AV-profiles for Sa
galaxies with small bulges agree better with what was published in GD15 than
for the ones with large bulges. The steeper \AV-profile slopes are the result
of the red-bulge population in their galaxy centers, which increases the
inferred extinction values. Before using the \betaV-method for spiral galaxies
with large bulges, these larger gradients need to be taken into account.

  %
\begin{figure}
\center
\includegraphics[width=0.489\txw]{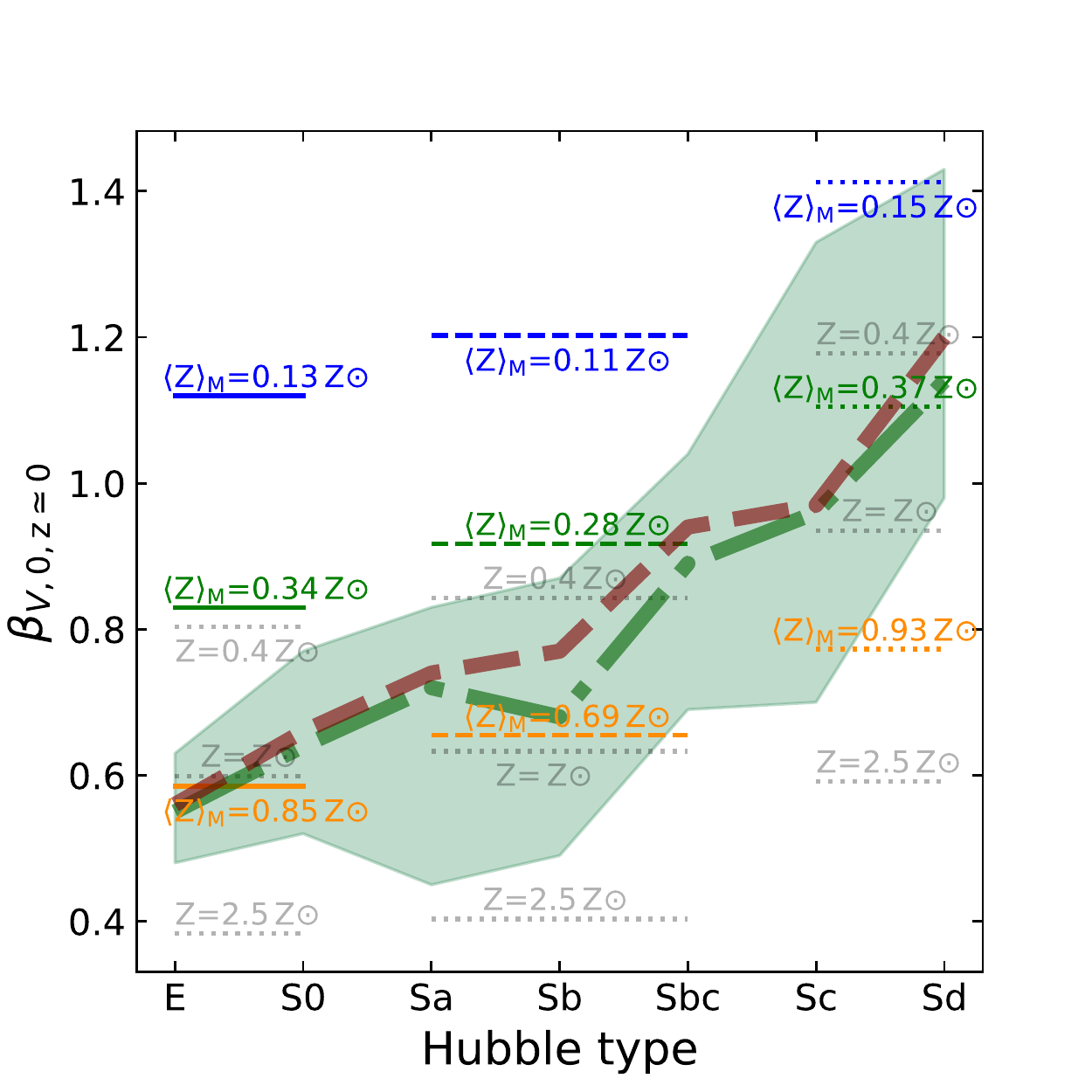}
\caption{\small
\betaVzero-values at $z$\,$\simeq$\,0 for galaxies of Hubble type E--Sd. The
green dot-dashed line and shaded region indicate the median and 1-$\sigma$
limits of the \betaVzero-values for galaxies in each Hubble type bin (see
Figure~\ref{fig:betav0_dist}). The thick brown dashed line indicates the
\betaVzero\ values derived from the median combined
$\langle$\betaV$\rangle$-profiles for each Hubble type. Horizontal lines show
the expected model values from K17, where the colors and line styles correspond
to those from Figure~\ref{fig:Zevol} (see Appendix D). Mass-weighted
metallicities for models at a redshift of $z$\,$=$\,0 are shown above or below
each line. \betaVzero-values with no metallicity evolution are plotted as light
gray dotted lines. Mass-weighted metallicity ranges for each Hubble type from
GD15 (see their Fig.11) match well to the mass-weighted metallicity range,
where \betaVzero-values are distributed throughout the green shaded region.
\label{fig:betav0}}
\end{figure}
  %

\item \emph{The K17 model holds for nearby galaxies.} K17 published the
intrinsic flux ratios of various visible -- near-infrared filters
(\betalamzero) with respect to the $L$ filter ($\sim$3.5\micron) as a function
of redshift for galaxies having SFHs characteristic of early-types (E and S0),
spirals (Sa--Sbc), and late-types (Sbc--Sd). K17 approximated metallicity
evolution by stacking the SEDs of simple stellar populations as a function of
SFH, which is different for each SFH type (see Appendix D for details).
Figure~\ref{fig:betav0} shows the \betaVzero-values at a redshift of
$z$\,$\simeq$\,0 from the K17 model, as well as the observed data from the
current study. The K17 model values are indicated by colored lines for
early-types, spirals, and late-types with metallicity offsets:
$Z(z$$=$$0)$\,$=$\,2.5, 1.0, and 0.4\,$Z_\odot$ from bottom to top. K17
selected the metallicity offset values arbitrarily, matching the metallicities
of empirical SED model of SSP. Metalliticy offsets result in mass-weighted
metallicities, $\langle$$Z$$\rangle$$_m$ at $z$\,$=$\,0, as indicated above or
below each line (see Figure~\ref{fig:Zevol}). The gray dotted lines are
\betaVzero-values without any metallicity evolution taken into account. The
\betaVzero-values derived in this study through \AV-profile matching are
plotted on top of the horizontal lines from K17. The thick green dashed line
and the corresponding green shaded region represent the median and 1-$\sigma$
ranges of the \betaVzero-values derived from the \betaV-profiles of individual
galaxies in each Hubble type bin, respectively (see
Figure~\ref{fig:betav0_dist}). The thick brown dashed line shows the
\betaVzero-values derived from the average \betaV-profile of all galaxies in
each Hubble type bin. The \betaVzero-values fall within a range of observed
$\langle$$Z$$\rangle$$_m$ from GD15 of 1.3$^{+0.7}_{-0.5}$,
1.2$^{+0.6}_{-0.4}$, 0.9$^{+0.5}_{-0.4}$, 0.8$^{+0.4}_{-0.3}$,
0.6$^{+0.4}_{-0.2}$, 0.4$^{+0.3}_{-0.2}$, and 0.3$^{+0.3}_{-0.1}$\,$Z\odot$ for
the E, S0, Sa, Sb, Sbc, Sc, and Sd types, respectively (see Figure~11 of GD15).

\end{enumerate}

The \betaVzero-values from the K17 model were confirmed by matching the
\betaVzero-values at a redshift of $z$\,$\simeq$\,0 to the observations. K17
derived \betaVzero-values using empirical models that were developed based upon
observations of galaxies including a sample at redshifts $z$\,$\gtrsim$\,2. The
K17 model at a redshift of $z$\,$\simeq$\,0 is the end-product of an
accumulation of SEDs of simple stellar populations as a function of SFH and
metallicity evolution. Any discrepancy at the intermediate redshift would have
resulted in the mismatch at redshift $z$\,$\simeq$\,0. Therefore, with \HST\
and future \JWST\ data, we expect the \betaV-method would work for galaxies at
a redshift $z$\,$\lesssim$\,2 with a careful classification of galaxy
morphology and metallicities.

\emph{JWST}/NIRCam's images are Nyquist sampled at 4\,\micron\ with
0.0647\arcsec\ pixels in the long wavelength arm \citep{Beichman12}. That
scale corresponds to 123, 220, and 406\,pc at redshifts of $z$\,$=$\,0.1, 0.2, and 0.5, respectively. NIRCam's pixel scale corresponds to a
physical scale larger than $\sim$200\,pc at redshift $z$\,$\gtrsim$\,0.2.
The same galaxies at these redshifts observed with both \HST\ and \JWST\ would
look similar to galaxies observed at a redshift of $z$\,$=$\,0.02 by \Spitzer\
at IRAC's 3.6\,\micron\ resolution, such as NGC\,1016, IC\,2239, NGC\,2832,
NGC\,2892, and NGC\,2937 from our sample (see the last paragraph of \S\ 4.1).
The spatial resolution effects from using the \betaV\ method in an \HST\ and
\JWST\ galaxy survey would therefore not be significant over the redshift range
0.2\,$<$\,$z$\,$<$2.0. For comparison, these spatial resolution effects
were evident in the \Spitzer\ observations in our sample at a redshift of
$z$\,$<$\,0.02. As a result of the increased resolution of \HST\ and \JWST,
regions smaller than giant molecular clouds would be resolved at redshifts of
$z$\,$<$\,0.2 \citep[$\lesssim$200\,pc]{Murray11}. Therefore the
\betaV-method using the average color over multiple stellar populations would
become less effective at these redshifts.

In addition to dust correction via the \betaV-map, the \betaV-method may also
serve as a detection tool for AGN and bluer/redder coherent features. Since AGN
are typically surrounded by thick dust tori, central \betaV-values lower than
the rest of the galaxy could serve as an indication of AGN activity. This
method was able to find previously known features, such as resonance rings and
\ion{H}{2} regions, and also an outflow-like feature that has not been observed
before.

\section{Summary}

Studying PSF-matched images of 257 NGC/IC galaxy mosaics in the SDSS $g$
and $r$ filters and \Spitzer\ 3.6\,\micron\ mosaics, we conclude the following:

\begin{enumerate}[label=\arabic*)]

\item The \betaVzero-values derived from \AV-profile matching with the
IFU-SED fitting results agree with the models from K17 to within the stated
errors.

\item \betaV\ becomes insensitive to small-scale variations in stellar populations once resolution elements subtend an angle larger than that of a typical giant molecular cloud ($\sim$200\,pc). At the resolution of our images this corresponds to a redshift $z$\,$\gtrsim$\,0.02; when combining \HST\ and \JWST\ images the method should be robust for $z$\,$\gtrsim$\,0.2.

\item SF/AGN galaxies have lower \betaV-values in the central regions.

\end{enumerate}

We conclude that the \betaV-method can serve as a simple dust correction method
for large galaxy surveys in the redshift range 0.2\,$<$\,$z$\,$<$\,2,
where \JWST\ observations will sample rest-frame $\sim$3.5\,\micron, and \HST\
observations sample rest-frame visible wavelengths. This will become
particularly useful when no, or only a limited number of, other broadband
images are available.

\vspace{1cm}
{Acknowledgements: This work is funded by NASA/ADAP grant NNX12AE47G (PI:
R.\,A.~Jansen). RAW acknowledges support from NASA JWST grants NAG5-12460,
NNX14AN10G, and 80NSSC18K0200. We thank Rosa M. Gonz{\'a}lez Delgado and
Rub{\'e}n Garc{\'{\i}}a-Benito for sending us the data of the plots in their
paper (GD15). DK thanks Wolfgang Steinicke for his permission to use his
Revised NGC and IC catalog. DK thanks Russell Ryan for sharing the \code{IDL}
code \code{segeditor} \citep{Segeditor}, which was very helpful for editing
\SExtractor\ segmentation maps. We appreciate the anonymous referee who
provided constructive comments and insights. We thank Bhavin Joshi, Teresa
Ashcraft, and Brent Smith for helpful discussion and comments. We thank Tom
Tyburczy and Garrett Rand for their help with coding PSF Extraction and
documenting the procedure. We thank Gabriela Huckabee for reviewing the draft
of this paper and giving helpful comments.

This work is based on observations made with the \Spitzer\ Space Telescope,
which is operated by the Jet Propulsion Laboratory, California Institute of
Technology under a contract with NASA. SEIP are provided by The \Spitzer
Science Center (SSC) and the Infrared Science Archive (IRSA).

Funding for the Sloan Digital Sky Survey IV has been provided by the Alfred P. Sloan Foundation, the U.S. Department of Energy Office of Science, and the Participating Institutions. SDSS-IV acknowledges
support and resources from the Center for High-Performance Computing at
the University of Utah. The SDSS web site is www.sdss.org.

SDSS-IV is managed by the Astrophysical Research Consortium for the 
Participating Institutions of the SDSS Collaboration including the 
Brazilian Participation Group, the Carnegie Institution for Science, 
Carnegie Mellon University, the Chilean Participation Group, the French Participation Group, Harvard-Smithsonian Center for Astrophysics, 
Instituto de Astrof\'isica de Canarias, The Johns Hopkins University, 
Kavli Institute for the Physics and Mathematics of the Universe (IPMU) / 
University of Tokyo, the Korean Participation Group, Lawrence Berkeley National Laboratory, 
Leibniz Institut f\"ur Astrophysik Potsdam (AIP),  
Max-Planck-Institut f\"ur Astronomie (MPIA Heidelberg), 
Max-Planck-Institut f\"ur Astrophysik (MPA Garching), 
Max-Planck-Institut f\"ur Extraterrestrische Physik (MPE), 
National Astronomical Observatories of China, New Mexico State University, 
New York University, University of Notre Dame, 
Observat\'ario Nacional / MCTI, The Ohio State University, 
Pennsylvania State University, Shanghai Astronomical Observatory, 
United Kingdom Participation Group,
Universidad Nacional Aut\'onoma de M\'exico, University of Arizona, 
University of Colorado Boulder, University of Oxford, University of Portsmouth, 
University of Utah, University of Virginia, University of Washington, University of Wisconsin, 
Vanderbilt University, and Yale University.

This research has made use of the NASA/IPAC Extragalactic Database (NED), which is operated by the Jet Propulsion Laboratory, California Institute of Technology, under contract with the National Aeronautics and Space Administration.

\code{PyRAF} is a product of the Space Telescope Science Institute, which is
operated by AURA for NASA.

}

\facilities{IRSA, NED, Sloan, Spitzer} 

\software{\code{Python}, \code{Numpy}, \code{Scipy}, \code{PyRAF},
\code{Astropy}, \code{IRAF}, \code{DS9}, \code{GALFIT}, \code{IDL},
\SExtractor; \code{segeditor} \code{IDL} code written by Russell Ryan.}

\bibliography{ms}

\appendix
\twocolumngrid

\section{PSF matching}

One needs to match PSFs of the images from various facilities before performing
any pixel-to-pixel analysis. A \code{Python}
script\footnote{\url{https://github.com/DuhoKim/PSFtractor}} for modeling PSFs
in the SDSS and \Spitzer\ mosaics was written to perform the PSF matching.
First, \SExtractor\ was run on each mosaic to generate a background image, as
well as a catalog of sources. These sources were then used as the input values
of the \code{PyRAF} \code{DAOPHOT} package \citep{DAOphot}. Stars having
\SExtractor\ parameter values listed below were selected and used to model the
image PSF with the \code{PyRAF} task \code{PSF}:
\begin{itemize}[noitemsep]
\item FLAGS\,$=$\,0
\item CLASS\_STAR\,$>$\,0.8 (0.7 for \Spitzer)
\item ELLIPTICITY\,$<$\,0.1
\item mode(FWHM\_IMAGE)\,$-$\,$\sigma$(FWHM\_IMAGE) \\
\hspace{2cm} $<$\,FWHM\_IMAGE\,$<$ \\
mode(FWHM\_IMAGE)\,$+$\,$\sigma$(FWHM\_IMAGE).
\end{itemize}
A radius of 30 pixels was used for the PSF model, which was large enough to
sample the wings of the PSF and small enough to exclude background objects. The
generated PSF model was then fed into the task \code {SUBSTAR}, which subtracts
any stars neighboring the PSF from the stars used to create the PSF-model. This
cleans out the background and models the PSF to a higher degree of accuracy.
This procedure was repeated one more time to finalize the modeling of each PSF.

PSF models of the \Spitzer\ mosaics were consistent in shape, but with varying
position angles (PAs). The PSF models were rotated to match the PAs, and then
stacked to build a master PSF for the \Spitzer\ mosaics. Before matching the
PSFs of the SDSS mosaics, the PSFs needed to be rotated back to the original
PA. The SDSS PSF model was constructed separately for each mosaic.

The \code{PyRAF} task \code{PSFMATCH} was used to match the PSFs of the SDSS
$g$ and $r$ mosaics to \Spitzer\ 3.6\,\micron\ mosaics. \code{PSFMATCH} convolved
the input mosaics using a convolution kernel $k=\mathscr{F}[R/I]$, where
$\mathscr{F}$ is the Fourier transform, $R$ is the Fourier transform of the
reference PSF (3.6\,\micron), and $I$ is the Fourier transform of the input PSF
(SDSS $g$ or $r$\,band) \citep{Phillips95}. Nearly identical results were found
regardless of the order that the tasks \code{WREGISTER}, \code{PSF}, and
\code{PSFMATCH} were performed in.

\section{Petrosian Half-light radius}

The half-light radius (HLR) was measured to normalize the radial profiles of
the galaxies so that they could be compared to those from the literature.

GD15 collapsed the spectral cubes in the rest-frame window 5635$\pm$45\,\AA,
which is near the central wavelength of the $V$ filter (5448\,\AA, see the
Table~7 of K17) and then measured the HLR. Their HLR values thus measured are
close to the Petrosian half-light radius
\citep[$R_{50}^{P}$;][]{Blanton01,Yasuda01} from the SDSS data archive
\citep[see][Appendix A]{GD14}.

To be consistent, we followed the definition of the SDSS Petrosian half-light
radius, $R_{50}^{P}$. A radial surface profile was first determined using the
median values of circular annuli in a radius range from 1 pixel to the
semi-major axis of the galaxy.

Next, the Petrosian radius, $R^P$, was calculated such that the flux within an
annulus between 0.8$R^P$ and 1.25$R^P$ was less than 20\% of the total flux
within the radius $R^P$. The radius containing 50\% of the flux within $R^P$
was defined to be the Petrosian half-light radius, $R_{50}^{P}$.
  %
\begin{figure*}
\centerline{
\parbox[b][0.355\txh][t]{0.015\txw}{\textbf{(a)}}
  \includegraphics[width=0.475\txw]{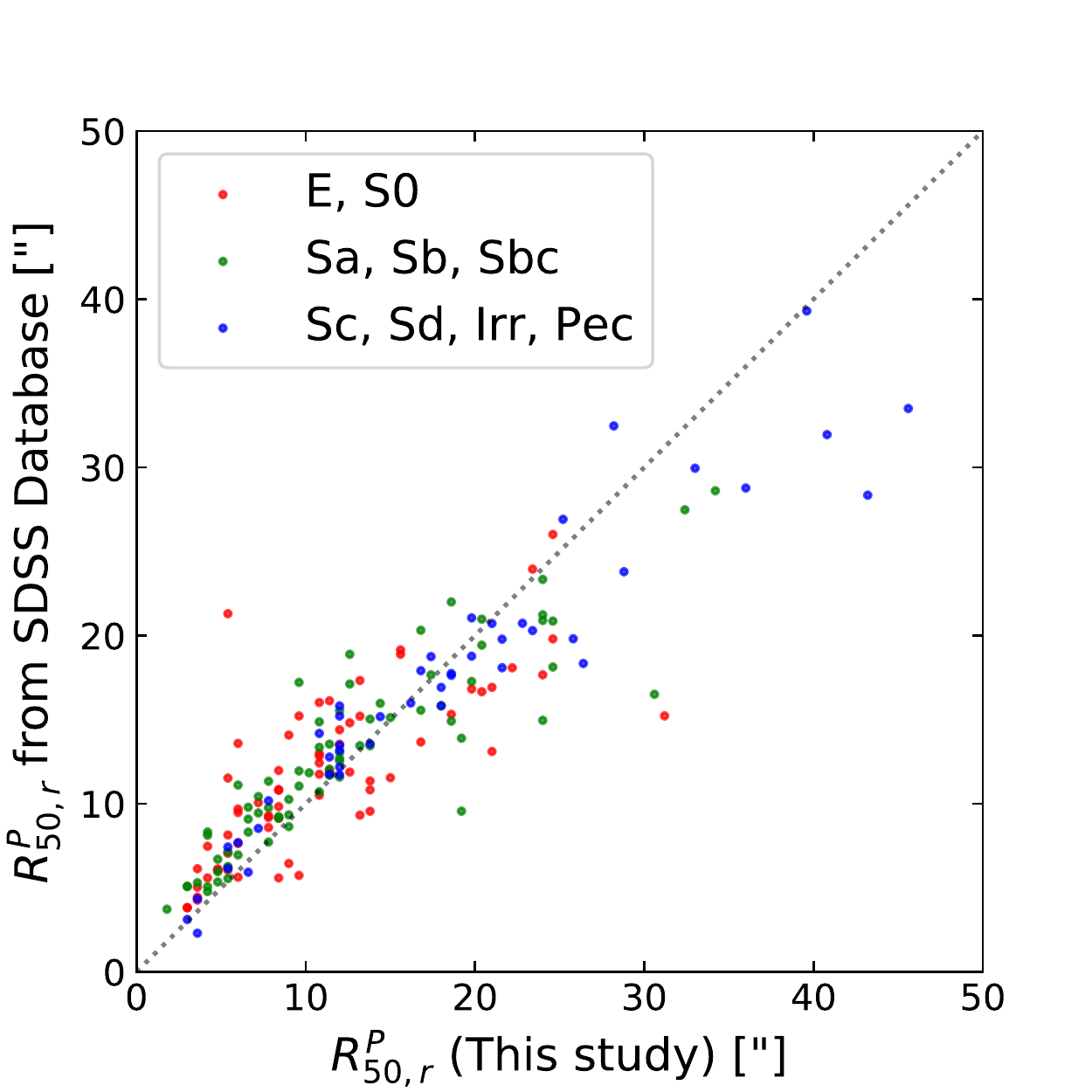}\hspace*{0.025\txw}
\parbox[b][0.355\txh][t]{0.015\txw}{\textbf{(b)}}
  \includegraphics[width=0.475\txw]{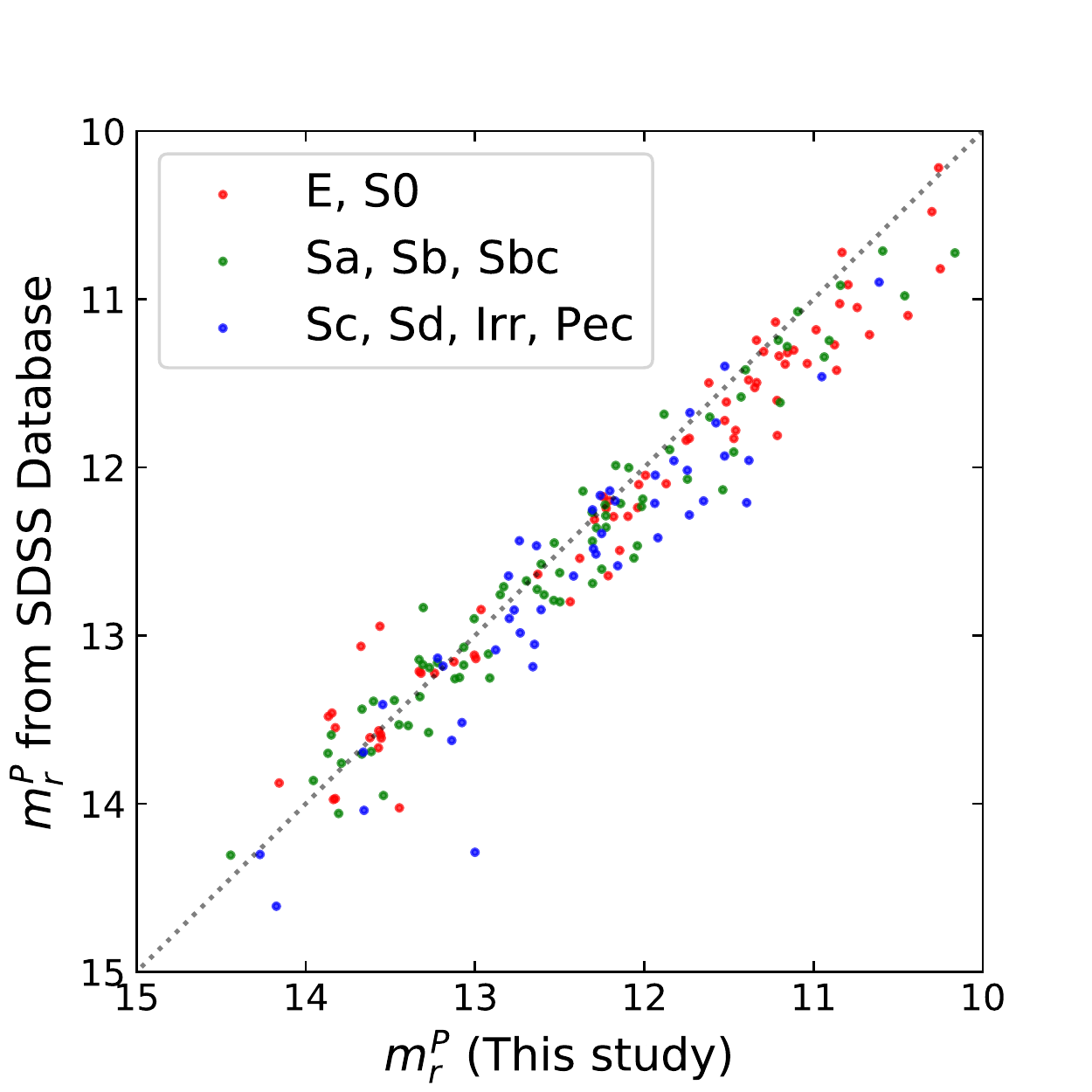}
}
\caption{\small
Comparison between (a) Petrosian half-light radii (\emph{petroR50\_r}) and (b)
Petrosian magnitudes (\emph{petroMag\_r}) in $r$\,band from the catalog `Galaxy'
in the SDSS archive and our measurements using downloaded SDSS mosaics. The
morphological types of galaxies are color coded. The almost one-to-one slope of
our measurements and the SDSS measurements shows agreement with the archival
catalog.\label{fig:HLR}}
\end{figure*}
  %
Figure~\ref{fig:HLR} shows the comparison between our measurements and the SDSS
data archive values of $R_{50}^{P}$ in $r$ (`petroR50\_r') of 182
coordinate-matched galaxies in our sample. The values from the `\emph{Galaxy}'
catalog on SDSS SkyServer
DR14\footnote{\label{sdss_cat}\url{https://skyserver.sdss.org/dr14/en/tools/cros
 sid/crossid .aspx}} were queried using the coordinates of our galaxy sample
with a 0.5\arcmin\ search radius. Overall, the galaxies match, so the
$R_{50}^{P}$ in the $V$\,band was used for the normalization of the radius in
Figure~\ref{fig:prof}.

\section{Bulge-to-total light ratio}

\code{GALFIT} \citep[version 3;][]{GALFIT1,GALFIT2} was used to measure the
bulge-to-total light ratios (B/T) of the sample galaxies. The segmentation map
generated for the \betaV\ analysis was used for the `Bad pixel mask'. The
3.6\,\micron\ mosaics for each galaxy, the master PSF of the 3.6\,\micron\
mosaics, and the `Bad pixel mask' were used as inputs for \code{GALFIT}. The
`Image region' and `Size of the convolution box' were set to the same as the
images shown in Figure Set~\ref{fig:figset2}), which is twice the semi-major
axis (column 7 in Table~\ref{tab:sample}).

Three components, the disk `expdisk', the bulge `devauc', and `sky', were input
to the fitting procedure to decompose the galaxy light into light from the
disk, the bulge, and the background, respectively. When a bright central point
source was found and the galaxy is known to have an embedded AGN (column 18 and
19 in Table~\ref{tab:sample}), we allowed for an additional `psf' component and
excluded that in the calculation of the bulge-to-total light ratio.

The following were used as the initial guesses for the \code{GALFIT} parameters: 
\begin{itemize}[noitemsep]
\item `position'\,$=$\,the physical pixel position converted
from R.A. and Decl.
\item `Integrated magnitude'\,$=$\,the magnitude in $V$
\item `R\_e'\,$=$\,one fifth of the semi-major axis
\item `b/a'\,$=$\,from Table~\ref{tab:sample},
\item `PA'\,$=$\,from Table~\ref{tab:sample}.
\end{itemize}
These parameters were then set as free, so they could be varied during the
fitting process. We chose one fifth of the semi-major axis as the initial guess
of the effective radius `R\_e'.

Upon visual inspection, any parameters that seemed unreasonable were fixed, and
\code{GALFIT} was rerun. For the case when either the bulge or disk moved
towards the outer part of the galaxy, the `position' argument of \code{GALFIT}
was set as fixed for both components. For the case where the bulge parameter,
`devauc', was seen to dominate the \code{GALFIT} fitting, a smaller effective
radius, `R\_e', was forcefully set for the `devauc' component to prevent
unphysical fitting at large `R\_e'. This was seen to be the case for some
late-type galaxies in the sample.

\begin{figure*}
\center
\includegraphics[width=0.9\txw]{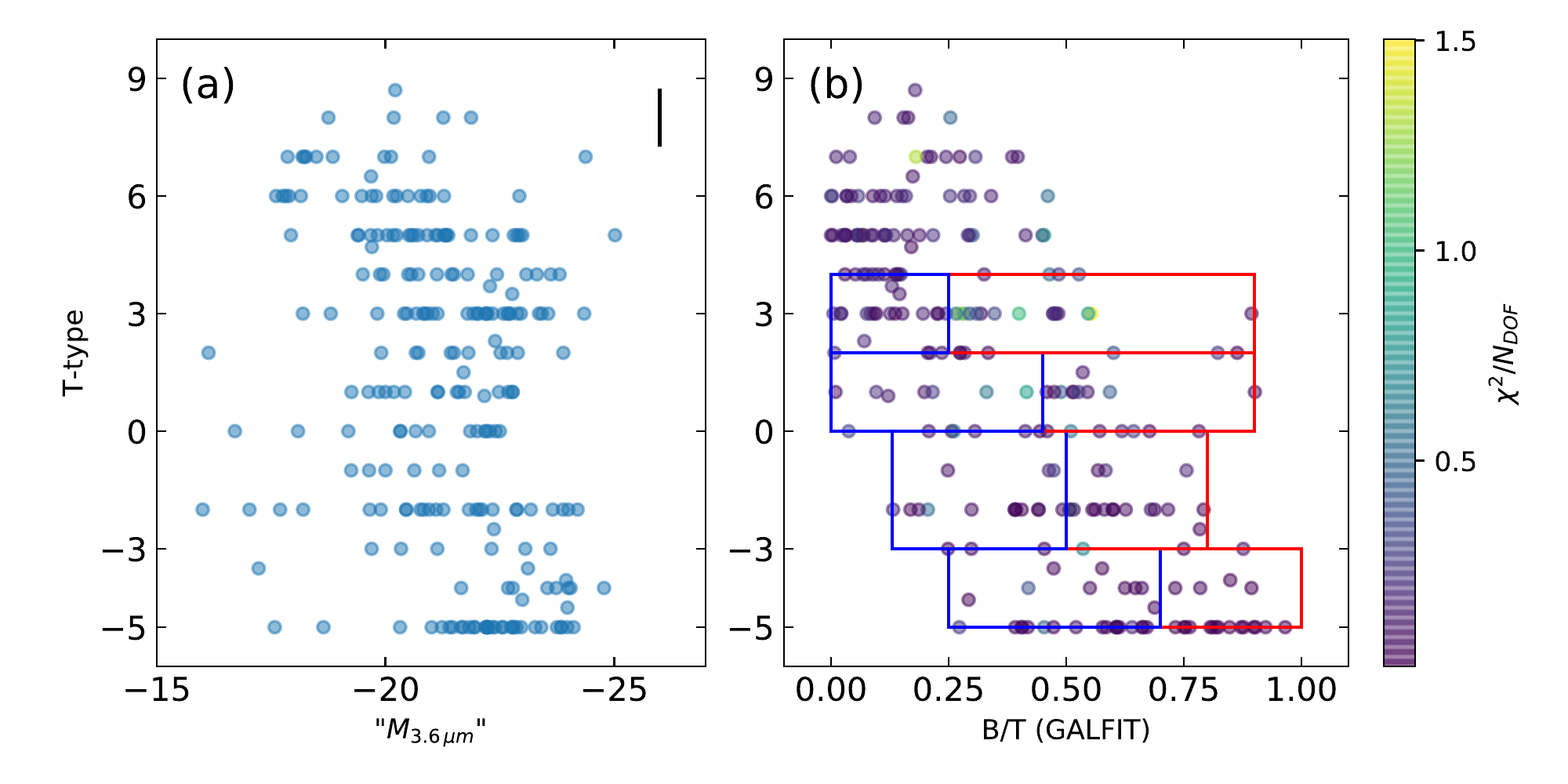}
\caption{\small
(a) RC3 T-type vs. total magnitude + distance modulus; (b) The bulge-to-total
light ratios (B/T) measured using \code{GALFIT} versus RC3 T-type (excluding
irregular and peculiar galaxies) of 239 galaxies. The colors of the dots in (b)
represent the $\chi^{2}$ divided by the number of pixels. The red (blue) boxes
contain galaxies whose B/T values higher (lower) than 0.7, 0.5, 0.45, and 0.25
for E, S0, Sa, and Sb type galaxies, respectively, which consists of the `Large
bulge' (red) and `Small bulge' (blue) subgroups for each Hubble types in
Figure~\ref{fig:prof_bt}.\label{fig:btot}}
\end{figure*}

Figure~\ref{fig:btot} shows the ``absolute magnitudes'' (a) and the B/T (b) as
a function of T-types in the RC3 catalog, excluding irregular and peculiar
galaxies (T\,$\geq$\,9). The `Integrated magnitude' output from the
\code{GALFIT} fitting for each of the `expdisk' and `devauc' components was
converted from magnitudes to flux units. The summed flux values were converted
back to the total magnitude of the entire galaxy. The distance modulus, which
was determined from the redshift, was added to the total magnitude of the
entire galaxy to determine the ``absolute magnitude'' for each galaxy. The
reduced $\chi^{2}$ value of each B/T fit is shown with different colors. The
bulge component, `devauc', dominates in the early-type galaxies (T\,$<$\,0),
while the disk component, `expdisk', gradually increases with T-type. The B/T
was used to subdivide the E--Sb galaxies in Figure~\ref{fig:prof}, which show
different shapes of \AV-profiles between this study and IFU-SED fitting method
used in GD15. E, S0, Sa, and Sb galaxies were divided using boundary B/T values
of $\simeq$ 0.7, 0.5, 0.45, and 0.25, respectively. Each group is shown in
Figure~\ref{fig:btot} (b) with colored boxes for galaxies with larger (red) and
smaller (blue) bulges than the dividing B/T values.
  %

\section{Mass-weighted Metallicity}

\begin{figure*}
\center
\includegraphics[width=0.65\txw]{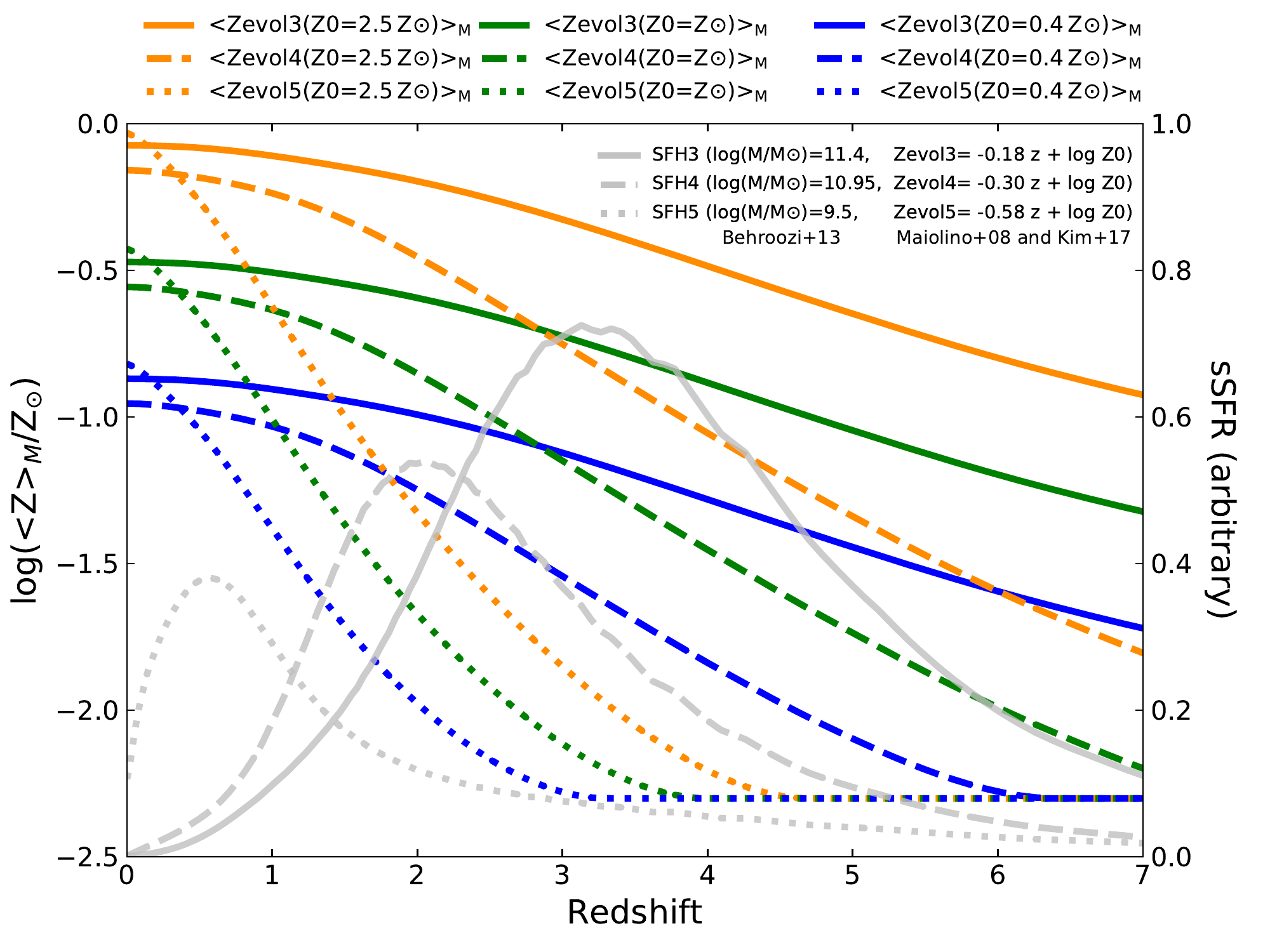}
\caption{\small
The mass-weighted metallicities, $\langle Z \rangle_\mathrm{m}$, and specific
SFRs (sSFRs) as a function of redshift from K17 are indicated by the colored
and gray lines, respectively. For sSFRs, K17 adopted SFRs of galaxies with
stellar masses of 10$^{11.4}$ (SFH3), 10$^{10.95}$ (SFH4), and
10$^{9.5}$\,M$\odot$ (SFH5) from \citet{Behroozi13}. Linear metallicity
evolution as a function of redshift was derived by K17 from these SFRs and the
mass-metallicity relation of \citet{Maiolino08}. The $\langle Z
\rangle_\mathrm{m}$ at a certain redshift, $z_c$, is the sum of the product of
metallicity and SFR divided by the sum of the SFRs in the redshift range
$z$\,$>$\,$z_c$. The lower metallicity limit at $\log
Z/Z\odot$\,$\cong$\,$-$2.3 was set by the lowest metallicity available from SED
models of simple stellar populations from \sBC\ and \sSB\ \citep[for details,
see][]{K17}.
\label{fig:Zevol}}
\end{figure*}

K17 published \betaVzero-values for stellar populations having stochastic SFHs
with metallicity evolutions taken into account. Figure~\ref{fig:Zevol} shows
the specific SFRs (sSFRs), and the mass-weighted metallicity, $\langle Z
\rangle_\mathrm{m}$, as a function of redshift. The gray solid,
dashed, and dotted lines indicate amplitudes of multiple exponentially
declining star-formation episodes for early-type (SFH3), spiral (SFH4), and
late-type (SFH5) galaxies, respectively. The SFH3, SFH4, and SFH5 models are
from \citet{Behroozi13} for galaxies with stellar masses of 10$^{11.4}$,
10$^{10.95}$, and 10$^{9.5}$\,M$\odot$, respectively.
  %

The $\langle Z \rangle_\mathrm{m}$ value is the sum of the products of SFR and
metallicity divided by the sum of the SFRs up to a certain redshift
($z$\,$>$\,x; see Equation~9 in K17). The colored solid, dashed, and dotted
lines in Figure~\ref{fig:Zevol} are the $\langle Z \rangle_\mathrm{m}$ values
of galaxies with SFH3, SFH4, and SFH5, respectively. K17 derived slopes for
linear cosmic metallicity evolution as a function of redshift using the
mass-metallicity relation from \citet{Maiolino08} and three SFHs from
\citet{Behroozi13}, which are $-$0.18 (SFH3), $-$0.03 (SFH4), and $-$0.58
(SFH5) \citep[see also Fig.20 of][]{Maiolino18}. The orange, green, and blue
lines represent different metallicity offsets, $Z(z$\,$=$\,0), of 2.5, 1.0, and
0.4\,$Z\odot$, respectively. The lower limit of the metallicity evolution was
selected as $Z$\,$=$\,0.0001, which originated from the available SED models
\BC\ \citep{BC03} and \SB\ \citep{SB99} that K17 used for young and old stellar
populations. For example, the solid orange line represents the evolution of the
$\langle Z \rangle_\mathrm{m}$ of a galaxy having SFH3 with a slope of the
linear metallicity evolution of $-$0.18 (Zevol3) with a metallicity offset,
$Z(z$\,$=$\,0), of 2.5\,$Z\odot$. Due to massive galaxies having high
metallicities and dwarf galaxies having low metallicities on average,
Figure~\ref{fig:Zevol} would not show many galaxies with dotted orange or solid
blue lines. Nonetheless, K17 showed all these results for completeness.

Nine $\langle Z \rangle_\mathrm{m}$ values at $z$\,$=$\,0 are shown in
Figure~\ref{fig:betav0} either above or below each colored line with the same
color and line type as the corresponding colored lines in
Figure~\ref{fig:Zevol}.

\section{\betaV\ vs. axis ratio}

\begin{figure}
\center
\includegraphics[width=0.5\txw]{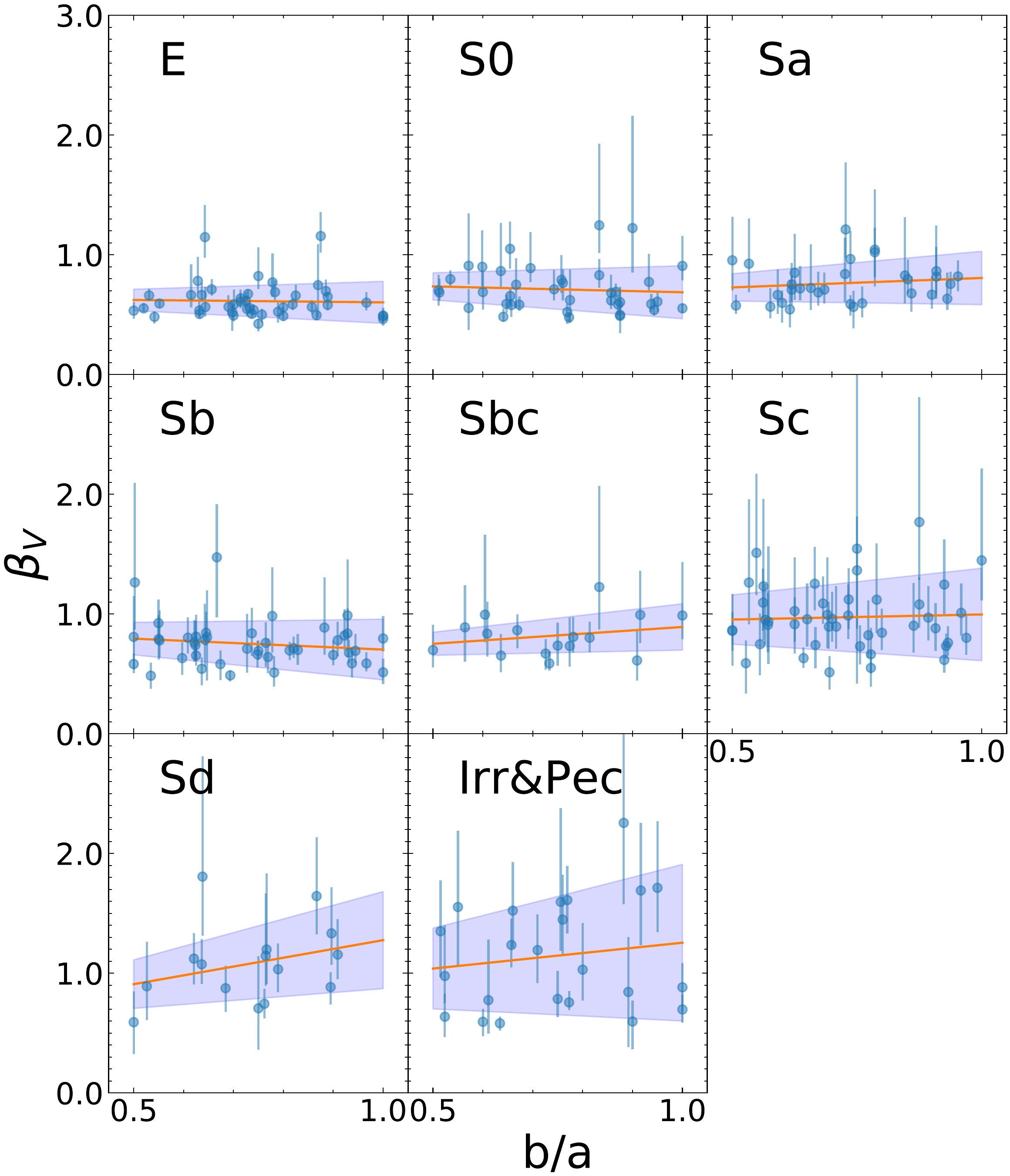}
\caption{\small
\betaV\ versus $b/a$ of individual galaxies for each in morphological type bin.
Galaxies with $b/a$\,$<$\,$0.5$ are not included in our sample. The best-fit
regression is overlaid as orange lines, while the shaded region represents the
$\pm$1-$\sigma$ range. Only Sbc, Sd, and Irr\&Pec galaxies in our sample show
hints of a correlation between the \betaV-value and the axis ratio for our
sample. However, within the uncertainties this correlation does not seem to be
statistically significant.\label{fig:incl}}
\end{figure}

To avoid potential systematic uncertainties originating from a wide range of
optical depths through edge-on galactic disks, we selected only face-on
galaxies having axis-ratios (b/a) larger than 0.5. Figure~\ref{fig:incl} shows
the relationship between \betaV-values and their axis-ratio values. No
significant correlation between \betaV\ and the axis-ratio was seen in our
sample to within the 1-$\sigma$ errors (blue shaded regions). Sbc, Sd, and
Irr\&Pec types may show at best a hint of a positive correlation, although this
is not significant within the current uncertainties.
  %

\end{document}